\documentclass[preprint,superscriptaddress,showpacs,preprintnumbers,amsmath,amssymb,prd]{revtex4-1}
\usepackage{graphicx}
\usepackage{dcolumn}
\usepackage{bm}

\newcommand{\mus}{\,$\mu $s}
\begin{document}

\title{Compton Scattering in Terrestrial Gamma-ray Flashes detected with the Fermi Gamma-ray Burst Monitor}

%
%
\author{Gerard Fitzpatrick}
\email{gerard.fitzpatrick@ucdconnect.ie}
\affiliation{School of Physics, University College Dublin, Belfield, Dublin 4, Ireland.}
\author{Eric Cramer}
\affiliation{Department of Physics and Space Sciences, Florida Institute of Technology, Melbourne, Florida, USA.}
\author{Sheila McBreen}
\affiliation{School of Physics, University College Dublin, Belfield, Dublin 4, Ireland.}
\author{Michael~S.~Briggs}
\affiliation{Center for Space Plasma and Aeronomic Research, University of Alabama in Huntsville, 320 Sparkman Drive, Huntsville, AL 35805, USA.}
\author{Suzanne Foley}
\author{David Tierney}
\affiliation{School of Physics, University College Dublin, Belfield, Dublin 4, Ireland.}
\author{Vandiver~L.~Chaplin}
\author{Valerie Connaughton}
\author{Matthew Stanbro}
\author{Shaolin~Xiong}
\affiliation{Center for Space Plasma and Aeronomic Research, University of Alabama in Huntsville, 320 Sparkman Drive, Huntsville, AL 35805, USA.}
\author{Joseph Dwyer}
\affiliation{Department of Physics and Space Sciences, Florida Institute of Technology, Melbourne, Florida, USA.}
\author{Gerald~J.~Fishman}
\affiliation{University of Alabama in Huntsville, Huntsville, AL 35812, USA.}
\author{Oliver~J.~Roberts}
\affiliation{School of Physics, University College Dublin, Belfield, Dublin 4, Ireland.}
\author{Andreas von Kienlin}
\affiliation{Max-Planck-Institut f\"{u}r extraterrestrische Physik, Giessenbachstrasse 1, 85748 Garching, Germany.}

\begin{abstract}
Terrestrial Gamma-ray Flashes (TGFs) are short intense flashes of gamma rays associated with lightning activity in thunderstorms. Using Monte Carlo simulations of the Relativistic Runaway Electron Avalanche (RREA) process, theoretical predictions for the temporal and spectral evolution of TGFs are compared to observations made with the Gamma-ray Burst Monitor (GBM) on board the \textit{Fermi} Gamma-ray Space Telescope. Assuming a single source altitude of 15\,km, a comparison of simulations to data is performed for a range of empirically chosen source electron variation timescales. The data exhibit a clear softening with increased source distance, in qualitative agreement with theoretical predictions.  The simulated spectra follow this trend in the data, but tend to underestimate the observed hardness. Such a discrepancy may imply that the basic RREA model is not sufficient. Alternatively, a TGF beam that is tilted with respect to the zenith could produce an evolution with source distance that is compatible with the data. Based on these results, we propose that the source electron distributions of TGFs observed by GBM vary on timescales of at least tens of microseconds, with an upper limit of $\sim$100\,$\mu $s.
\end{abstract}
\maketitle

\section{Introduction}
Terrestrial Gamma-ray Flashes (TGFs) are short intense flashes of gamma rays associated with lightning activity in thunderstorms which were discovered serendipitously in 1994 by the Burst And Transient Source Experiment (BATSE) \citep{Fishman1994}. TGFs are characterised by short timescales ($<1$\,ms) and hard spectra which can extend up to tens of MeV \citep{Smith2005}. Since their discovery, TGFs have been extensively studied by the Reuven Ramaty High Energy Solar Spectroscopic Imager (RHESSI) \citep{Grefenstette2009}, Astro‐rivelatore Gamma a Immagini LEggero (AGILE) \citep{Marisaldi2013}, and the Gamma-ray Burst Monitor (GBM) and Large Area Telescope (LAT) on board the \emph{Fermi} Gamma-ray Space Telescope \citep{Briggs2013,GROVE_REF}. The exact emission mechanism of TGFs is unknown, but the leading theoretical models involve the Relativistic Runaway Electron Avalanche (RREA) process, whereby electrons are accelerated to high energies in electric fields \citep{1992PhLA..165..463G}. As they propagate through the atmosphere, these electrons emit gamma rays via bremsstrahlung. The spectral and temporal properties of many averaged TGFs have been compared to RREA simulations and found to be broadly consistent for RHESSI observations, e.g. \citet{Smith2005,2008GeoRL..3506802G,Marisaldi2010a}. However, observations of a power law extending up to 100\,MeV by AGILE have challenged this view, as such a spectral shape is inconsistent with standard RREA models \citep{Tavani2011}.

The high count rate (on the order of hundreds of kHz) and low statistics associated with TGFs greatly complicates their analysis. The large effective area of the BATSE detectors allowed the study of TGFs on an individual basis, e.g. \citet{Feng2002,Nemiroff1997,Ostgaard2008}. However, these observations were later found to have been heavily modified by instrumental dead time \citep{Grefenstette2009,Gjesteland2010}. In general, RHESSI does not collect enough counts per TGF to study them on an individual basis. Consequently, analysis of this data has concentrated on stacking many TGFs and studying the average behaviour, e.g. \citet{Smith2005,Grefenstette2009}. The stacking of RHESSI events is necessary but unfortunate, as it combines many TGFs with differing orientations and source\,--\,detector geometries.  As the distance between the source of the TGF and the observer is increased, the entire TGF is expected to soften and be temporally extended, as a greater proportion of the collected counts have undergone Compton scattering \citep{Ostgaard2008,2011JGRA..11603315C,Celestin2012} due to the greater integrated depth of atmosphere traversed. 

TGFs typically consist of an individual pulse, but can be composed of multiple emission episodes. The time profile of a pulse can be either symmetric (comparable rise and fall time) or asymmetric, with a faster rise time than fall time \citep{Foley2014}. This temporal asymmetry and the tendency to soften in time was first noted in BATSE TGFs \citep{Nemiroff1997}. This softening was quantified as the temporal lag between the peak of soft (25\,-\,110\,keV) and hard ($>$\,110\,keV) counts in 15 BATSE TGFs \citep{Feng2002}. The lags ranged from $\sim$\,70\,-\,370\,$\mu s$, with an average value of $\sim$100\,$\mu s$. A detailed analysis of Monte Carlo simulations showed that these lags could be explained as being solely due to Compton scattering \citep{Ostgaard2008}. A study of 431 combined dead time corrected RHESSI TGFs found an average delay between the arrival times of the soft ($<$300\,keV) and hard counts to be 28\,$\pm$\,3\,$\mu s$ \citep{2008GeoRL..3506802G}. The discrepancy between the two values is due to the effects of dead time which were not accounted for in the initial studies of BATSE data.

Although not as sensitive as BATSE, GBM still accumulates enough counts per TGF that they can be studied on an individual basis. The implications of dead time on GBM observations are also less severe than BATSE. The absolute timing accuracy of GBM allows correlations of TGFs detected with individual lightning strokes, e.g. \citet{Connaughton2012}. For the subset of TGFs detected by GBM with associated radio detections, the source location and hence the distance and relative orientation of the spacecraft to the source can be determined. This allows a more detailed study of the temporal evolution of TGFs as a function of source to satellite distance than has previously been possible. Using the Runaway Electron Avalanche Model (REAM), a Monte Carlo code developed by J.~R.~Dwyer at Florida Institute of Technology to simulate the RREA model \citep{Dwyer2003,Dwyer2007}, we obtain the predicted temporal evolution as a function of source to satellite distance. However, these predictions cannot be directly compared to the observations of GBM, as the counts observed by an instrument have been distorted by the detection process. Additional distorting effects can arise due to instrumental effects such as dead time and pulse pile-up. 

In order to compare the model predictions to the observations, the predictions must be folded through the Detector Response Matrices (DRMs) and passed through a dead time filter. A GBM DRM is a lookup table which maps incident photons to detector counts. It is not diagonal, as incident photons do not necessarily have to deposit all their energy in the detector. 
The DRMs and dead time distort the incident photon distribution, with the result that varying source models may appear similar. This degeneracy is unfortunate, but unavoidable with the current generation of instruments. 

\section{Observations and Data}
The \textit{Fermi} Gamma-ray Space Telescope consists of two instruments, the Large Area Telescope (LAT) \citep{Atwood2009} and the Gamma-ray Burst Monitor (GBM) \citep{Meegan2009}. GBM consists of 14 individual scintillation detectors, 12 sodium iodide (NaI) with an effective energy range of 10-1000\,keV, and two bismuth germanate (BGO) crystals with an effective energy range of 0.2\,-\,40\,MeV.  The large effective area of the BGO detectors ($\approx 160$\,cm$^2$ per detector \citep{Tierney2013}) is particularly important as it facilitates the accumulation of sufficient statistics such that TGFs can be studied on an individual basis.

The relative timing resolution of 2\,$\mu $s and absolute timing accuracy of several microseconds, allows the correlation of gamma rays with individual lightning strokes. This is vital, as it allows the source location, and therefore the orientation and off-axis distance of the spacecraft relative to the source to be calculated.

In the period from the launch of \textit{Fermi} to March 16$^{\textrm{th}}$ 2013, GBM has detected $\sim$\,1330 TGFs. This includes TGFs which triggered GBM and those which have been found in an off-line search. A preliminary correlation study has confirmed radio matches with the World Wide Lightning Location Network (WWLLN) and/or the EARTH Networks Total Lightning Network (ENTLN) for 287 of these TGFs. For each TGF in the sample, the radio location was used to determine the distance and orientation of \textit{Fermi} to the source. Using this information, DRMs were generated for each individual TGF. The data used in this work are GBM Time-Tagged Events (TTE), which have a relative timing resolution of 2.0\,$\mu $s and 128 pseudo-logarithmically spaced energy channels.

\section{Simulations}
Comparing RREA simulations to individual TGFs is an extremely complex task, as the exact electric field (\textit{E}) orientation and beam geometry at the source is unknown. However, a statistical study in which we assume an altitude and beam geometry can be used to study the effect of increasing source offset from the satellite on the simulations. These simulations can be folded through the DRMs and dead time filtered to create `synthetic' TGFs. These can then be analysed in exactly the same fashion as the data. In this way, a direct comparison can be made between the theoretical predictions and the observed trends in the data.

\subsection{Simulating RREA time profiles}
To generate the predicted temporal and spectral distributions, we use the REAM Monte Carlo simulation code with an ambient DC E field value of 400 kV/m. An instantaneous electron distribution of the form
	\begin{equation}f(E)\propto\exp(-E/7.3\mbox{\,MeV})\end{equation}
is created at the source with $10^5$ electrons. In the presence of the \textit{E} field, these electrons undergo RREA multiplication. The electron interactions in the simulation include ionisation, atomic excitation and M\o{}ller scattering. Elastic scattering is fully modelled using a shielded Coulomb potential and includes bremsstrahlung production of X-rays and gamma rays. Compton scattering is also fully modelled and allows for the production of secondary electrons. After five avalanche lengths ($\lambda \approx 50$\,m), the resulting distribution of $\sim 1.6\times10^8$ photons is propagated from the source to an altitude of 100\,km. At this point, the atmospheric density is sufficiently low that the photons can be simply translated to 565\,km, the altitude of GBM.  These electrons travel along field lines and can also create more photons via bremsstrahlung.

\begin{figure}\center
	\includegraphics[]{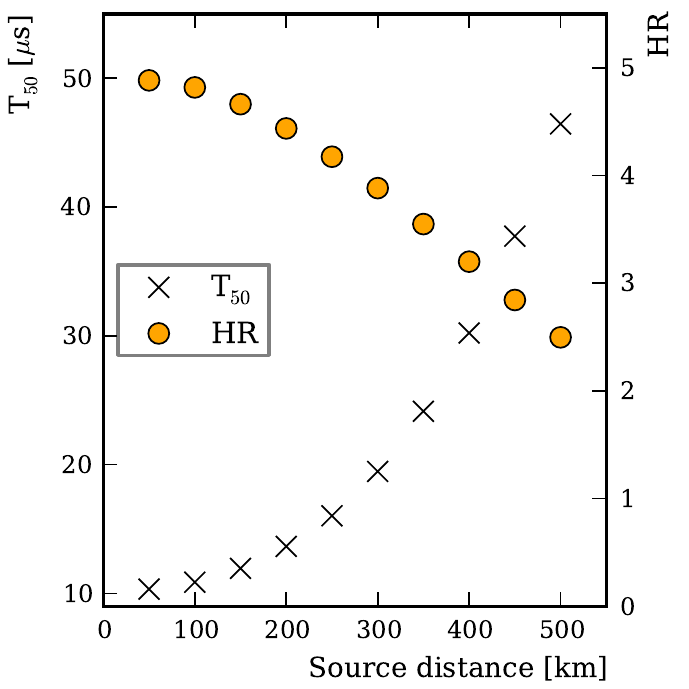}
	\caption{Evolution of the duration (crosses) and hardness ratio (circles) of the RREA simulations at an altitude of 565\,km, as a function of the TGF source to sub-satellite distance for a 15\,km altitude wide beam source model. The duration is measured using the $T_{50}$, the time interval in which 50\% of the flux occurs, starting and ending at 25\% and 75\% levels. The hardness ratio is given by the number of events with energy greater than 300\,keV divided by the number of events with energy less than 300\,keV. As the source off-axis distance is increased, a clear elongation in time and spectral softening is visible. This is due to the increased Compton scattering experienced by the photons as they propagate through a greater integrated density of atmosphere. }\label{f:sim}
\end{figure}

Once propagated to the spacecraft altitude, the photon distribution is integrated into concentric annuli of diameter 50\,km. This ensures that there are sufficient statistics in each ring while also allowing the effect of an increased source distance to be studied. We use a source altitude of 15\,km and a wide beam geometry, as this is believed to be typical for TGFs (e.g. \citet{Smith2005,2008GeoRL..3506802G,Carlson2007,Ostgaard2008,Hazelton2009,Gjesteland2011} ). For this geometry, the photons are spatially spread into a 45 degree isotropic cone (half-angle), simulating a diverging electric field at the source. Figure~\ref{f:sim} shows the evolution of the temporal and spectral properties as a function of source distance. In this Figure, the expected spectral softening and temporal elongation with increased distance is clearly evident.

\subsection{Synthetic TGFs}

To generate synthetic TGFs with RREA time profiles and spectra, we first consider a source directly below the satellite. This source is moved away from the satellite in 50\,km steps. For each step, 1000 synthetic TGFs are generated by randomly drawing a fluence of photons for each detector from the predicted temporal and spectral distributions for that offset. These are then folded through the DRMs and a dead time filter is applied. Poisson noise is added to each detector to simulate a background component. The resultant simulated TGFs have now undergone the same processes as the real data. 

To generate a synthetic TGF incident on GBM with a photon fluence of $n$ ph/cm$^2$, the following steps are taken. First, a set of DRMs is created using the known source location. Next, the appropriate annulus from which to draw photons is selected. For each GBM detector, the number of incident photons ($n_{\mbox{i}}$) is given by the product of $n$ and the geometric area of the detector ($A_{\mbox{g}}$), the projection of the surface area of the detector in the direction of the source location. The desired number of photons, $n_{\mbox{i}}$, is randomly drawn from the source distribution and folded through the DRM for that detector. 

To fold a single photon with a DRM, the DRM must first be converted to a probability distribution by dividing by $A_{\mbox{g}}$. The appropriate channel column in this matrix is then be selected. The sum of this column is the probability that the photon deposits any energy in the detector. A random number is drawn and if it is less than the summed probability, a random channel is drawn from the column; otherwise, the photon is discarded. The background is simulated as a homogeneous Poisson process, with a rate set to the mean of the observed background rate distribution for the NaI and BGO detectors, which is 1.1 and 1.6\,kHz respectively.  Each count is assigned a corresponding energy value which is randomly drawn from the observed background count spectrum distribution. The background is then combined with the source counts and passed through an instrumental filter that includes the effects of both dead time and pulse pile-up and is based on \citet{Chaplin2013}.

As the off-axis distance is increased, the photon fluence incident on GBM is normalised to match the observed counts. This is achieved by converting the observed count fluence to photons/cm$^2$ by dividing by the effective area. For this step, we consider only the observed fluence in the BGO detectors, as for these, the influence of dead time on the observed fluence has been studied in detail (e.g. \citet{Briggs2010,Tierney2013}). NaI detectors suffer from a greater effective dead time due to the higher proportion of overflow counts compared to the BGO detectors (overflow counts, which are above the maximum digitised energy, incur a 10\,$\mu $s dead time).

\begin{figure}\center
	\includegraphics[width=0.5\textwidth]{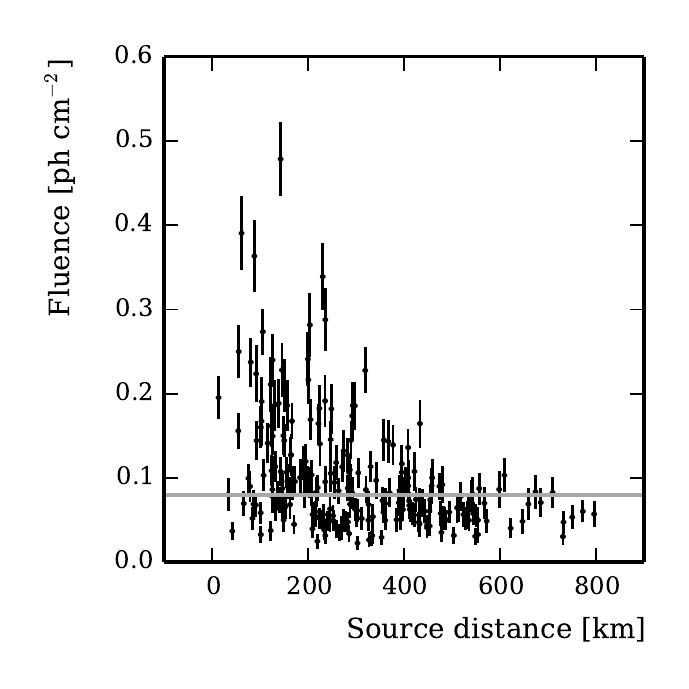}
	\caption{Photon fluence in BGO detectors as a function of the source to satellite distance. These values have not been corrected for dead time losses. The horizontal grey line represents the cut that was applied to ensure robust statistics for the analysis (see \S~\ref{s:analysis}).}\label{f:flu}
\end{figure}

For each real TGF in our sample, a 15\,km source model with a fluence of 1 ph/cm$^{2}$ was folded through the response of each BGO detector. The effective area is given by the ratio of the observed counts from the DRM-folded model to the number of incident photons. The incident photon fluence of each TGF is then calculated by dividing the number of observed counts in the data by the effective area. Applying this method to the entire sample allows the approximate source fluence in ph/cm$^{2}$ as a function of the TGF source to sub-satellite distance to be determined. This is shown in Figure~\ref{f:flu}. The fluence is binned into 50\,km bins to match the simulations and the average fluence in each bin found. In the absence of dead time, these values could be used as the incident photon fluence in the simulations. However, the dead time is non-negligible and a correction factor must be applied. 

The actual percentage loss depends on the incident flux, but losses of 60\,\% during the peak emission have been estimated for a very bright TGF \citep{Briggs2010}. For simplicity, we assume an average loss of 40\,\%. The corrected fluence in each 50\,km bin is used as the incident photon fluence in the corresponding simulation bin.

The REAM code generates an initial source electron distribution that is created instantaneously. In reality, the electron distribution will have a time dependence (e.g. \citet{Dwyer2012a,Celestin2012a}). To add this feature to our simulations, we artificially smear the simulated photon arrival times at the spacecraft with a Gaussian distribution. As there are limited predictions in the literature for the timescale of the variation at the source, we consider 4 empirical smearing distributions, with standard deviations ($\sigma$) of 25, 50, 75 and 100\,\,$\mu $s respectively. These values were selected to be representative of the durations observed at spacecraft altitudes.

\begin{figure}\center
	\includegraphics[width=7cm]{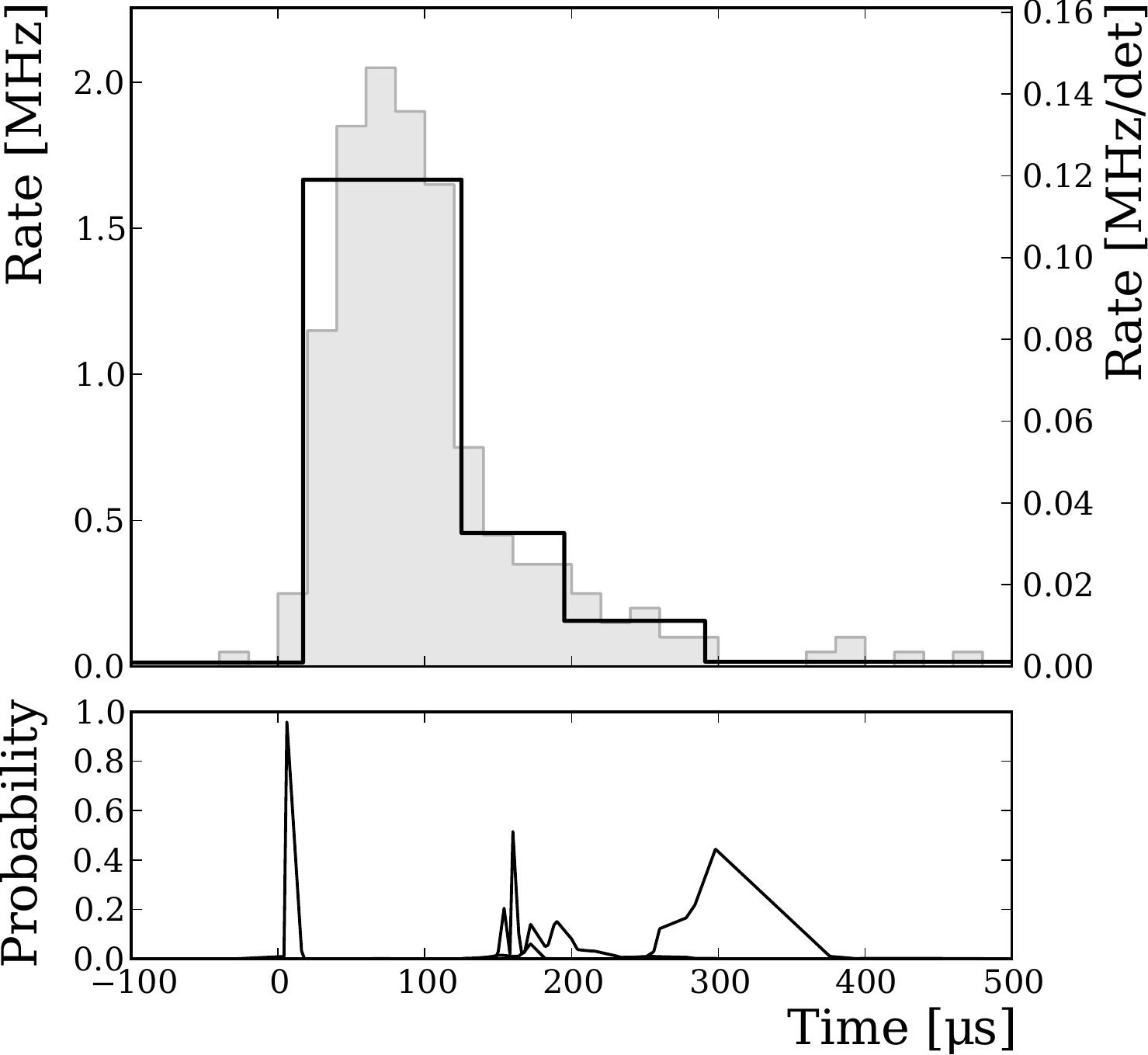}
	\caption{Results of the Bayesian Block Algorithm for TGF081123. The top panel shows the time profile, with the integrated counts in 20\,$\mu $s bins shown in grey. The solid line is the optimum representation of the data. The bottom panel shows the probability distribution for each change point. }\label{f:bba}
\end{figure}

\section{Analysis}\label{s:analysis}
In this work we analyse hundreds of real, and thousands of synthetic TGFs. It is desirable that we have an analysis method that is as objective and automatic as possible. The most problematic step is determining the point at which the TGF begins and ends.  Once the interval is chosen, auxiliary information such as the hardness ratio can then be calculated over this interval.

In previous analyses of GBM data, the $T_{90}$ measure has been used to define the time interval and duration of TGFs (e.g. \citet{Briggs2010}). Adopted from the study of GRBs, this takes the duration of an event as the time interval in which the fluence rises from 5\,\% to 95\,\% \citep{Koshut1996}. The time interval in which the counts are accumulated is defined by the user, and thus requires human interaction. Due to the low number of counts detected, and general trend for the intensity of TGFs to fall with time, determining the times at which the TGF is indistinguishable from instrumental background can be difficult.  Ref.~\citep{Connaughton2012}, uses the related $T_{50}$ measure as it is less susceptible to uncertainties caused by low count rates and background. However, the $T_{50}$ is not necessarily representative of the TGF duration. For these reasons, we do not use the $T_{90}$ or $T_{50}$ measures in this work.

To select the time intervals corresponding to the TGF for both actual observations and simulations, we employ the Bayesian Block Algorithm (BBA). This is a non-parametric algorithm that finds the optimal segmentation of data \citep{Scargle2013}. It is designed to address the general problem of detecting and characterising variability in time series data and can be applied to both time tagged and binned data. The data are divided into discrete segments or blocks, in  which the rate is modelled as a constant.  In practice, this translates to binning the data into non-uniform bins of common rate. This algorithm is frequently used in X-ray and gamma ray astrophysics (e.g. \citet{Buehler2012}). A brief overview of the algorithm follows.

The algorithm iterates over the data, adding in one data point with each iteration. As each data point is added, all possible segmentations of the data are tested. The segmentation which maximises the fitness is chosen, the expression for which depends on the data type. For GBM TTE data, the relevant expression is given by Eq.~19 in \citet{Scargle2013}. The number of blocks is not explicitly set, but is influenced by defining a prior distribution for the number of blocks. Ideally this prior should assign higher probability to a small number of blocks. The geometric prior, given by $P(N)=p \gamma ^{N_{\text{blocks}}}$, is used. As $N_{\text{blocks}}$ is not known in advance of the analysis, the contribution of the prior to the fitness ($\textit{ncp}_{\text{prior}}$) is introduced. Using simulations of pure-noise time series, $\textit{ncp}_{\text{prior}}$ is related to a false positive probability $p_0$. By adjusting $p_0$, the prior distribution is adjusted. A complete description of the algorithm can be found in \citet{Scargle2013}.

Using $p_0=0.05$, BBA was run for each TGF. The blocks corresponding to the TGF were selected by comparing them to the background rate. The time interval of these blocks was then taken as the duration of the TGF ($T_{BB}$). Figure~\ref{f:bba} shows an example of how the BBA technique is applied to a TGF.

\begin{figure}
	\centering
	\noindent\includegraphics{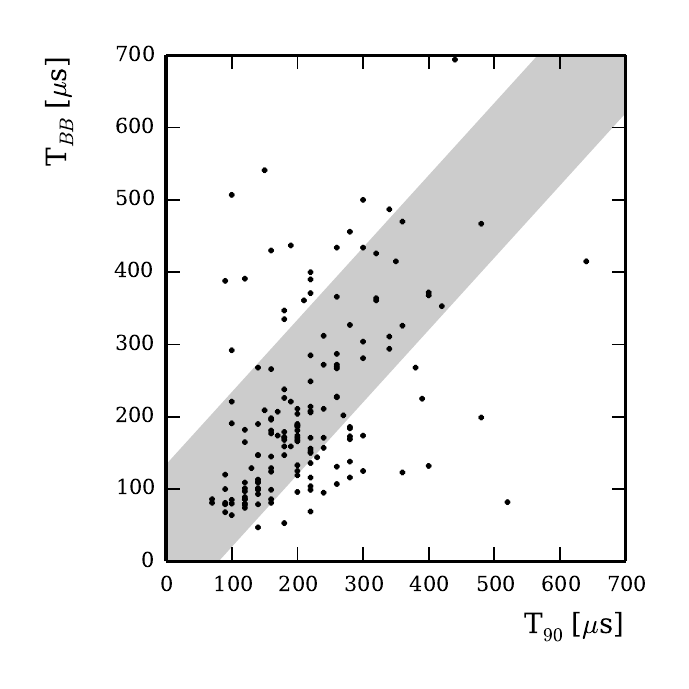}
	\caption{Comparison between duration calculated using Bayesian Blocks ($T_{BB}$) and $T_{90}$ values for the intersection of TGFs used in this work and \citet{Connaughton2012}. The shaded region indicates the 68\% containment region.
	}\label{f:comparion}
\end{figure}

In comparison to $T_{90}$, $T_{BB}$ is more conservative while also being less subjective. To compare the results from the Bayesian Block analysis to the $T_{90}$ measure, we use the intersection of the sample used in this work and that used in \citet{Connaughton2012}. For 158 common TGFs, the $T_{90}$ values are plotted against the corresponding $T_{BB}$ in Figure~\ref{f:comparion} (although only $T_{50}$ values were published in \citet{Connaughton2012}, $T_{90}$ values were also produced by the same analysis). To quantify the degree of correlation between the two measures, we use the Pearson product-moment correlation test, obtaining a coefficient of 0.47 which implies a moderate degree of linear correlation and indicates that the two measurements are broadly consistent.

For each TGF, the duration was taken as the time interval defined by $T_{BB}$. The hardness ratio (HR) of an event is defined as the ratio of counts above an energy threshold to those below it. This measure is useful as it can be used as an identifier for spectral evolution. However, the HR is heavily instrument dependent as it is based on count data, without any deconvolution to incident photons. Following previous studies (e.g. \citet{2008GeoRL..3506802G}), we have adopted the convention of defining soft counts as those with measured energies $<$\,300\,keV. For each TGF, the HR ($E_{>300}/E_{\le 300}$) was calculated over the time range $T_{BB}$.

Following \citet{2008GeoRL..3506802G}, we characterise the spectral evolution of a TGF by the delay between the counts above and below 300\,keV. This is calculated over the interval $T_{BB}$ by first finding the average arrival time of each component and then calculating the difference. The error on each component was taken as the standard error on the mean. The error on the delay is the quadrature sum of each.

As the source distance is increased there is a decrease in the observed fluence. To mitigate the effect of this on the analysis and to ensure robust statistics, TGFs with an observed fluence less than 0.08 ph cm$^{-2}$ and source distances greater than 500\,km were discarded. The fluence cut is shown as a horizontal grey line in Figure~\ref{f:flu}.

The simulated TGFs were analysed in an identical fashion to the data. To facilitate the comparison of this analysis to the data, the distributions of derived parameters (duration, delay, HR) were analysed and fit with a Gaussian for each set of 1000 simulations. The mean and standard deviation of each fit could then be compared to the data directly.

\section{Results}
\begin{figure}\center
	\noindent\includegraphics{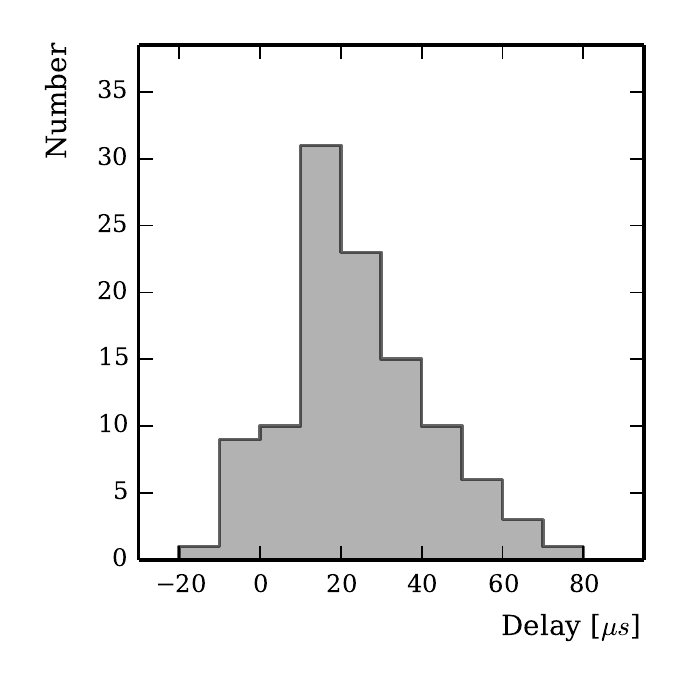}
	\caption{Distribution of the observed delay between soft and hard counts. }
\label{fig:delayDist} \end{figure}

The distribution of observed delays in TGFs is shown in Figure~\ref{fig:delayDist}. The observed values range from -20 to 80\,$\mu $s with a mean of $\approx24$\,$\mu $s. It is tempting to compare this to the value of 28\,$\mu $s obtained in the RHESSI analysis of 431 stacked TGFs \citep{2008GeoRL..3506802G}, however care must be taken as as the delays are calculated from counts detected in various energy bands, measurements that are detector dependent. The RHESSI data were also corrected for dead time losses. 

\begin{figure}\center
	\noindent\includegraphics[width=8cm]{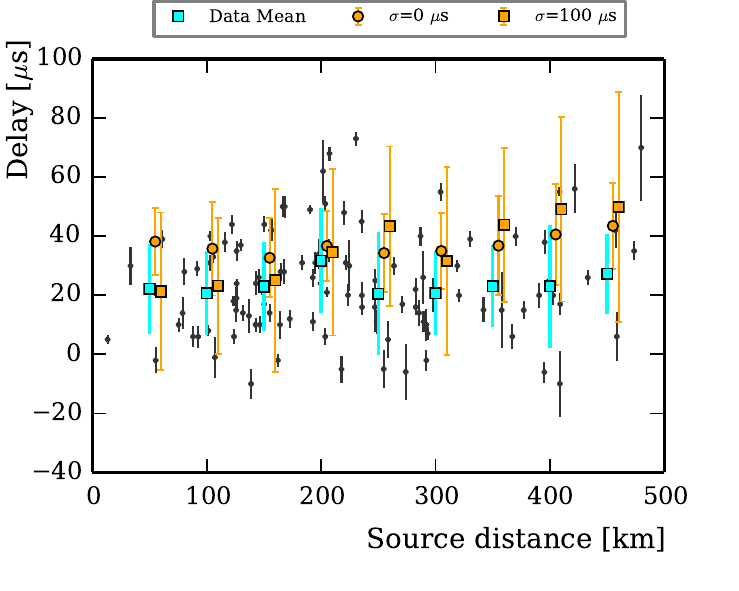}
	\caption{Delay between soft and hard counts as a function of the distance between the source and the \textit{Fermi} sub-satellite position. The individual TGFs are plotted in grey, and the average in 50\,km bins is plotted as cyan squares with the standard deviation plotted as an error bar. The values obtained in the simulations are plotted in orange. For the sake of clarity, the simulated values are offset from the data in 5\,km steps. To indicate the spread of the simulated values, the standard deviation of the fit is plotted as an error bar. The different smearings for the simulations are indicated by the marker, circles for no smearing and squares for 100\,$\mu $s. }
\label{fig:delay} \end{figure}

The delay as a function of the TGF source to sub-satellite location distance is shown in Figure~\ref{fig:delay} for the data and simulations. The data exhibit no significant variation with offset. The simulations which have undergone no smearing show no significant variation and are consistently longer than the mean of the data. The four electron timescales considered for the smeared simulations are all broadly consistent, and up to 200\,km, are consistent with the mean of the data. After this, similar to the zero smearing cases, the simulated delays are consistently longer than the data. 

\begin{figure}\center
	\noindent\includegraphics[width=8cm]{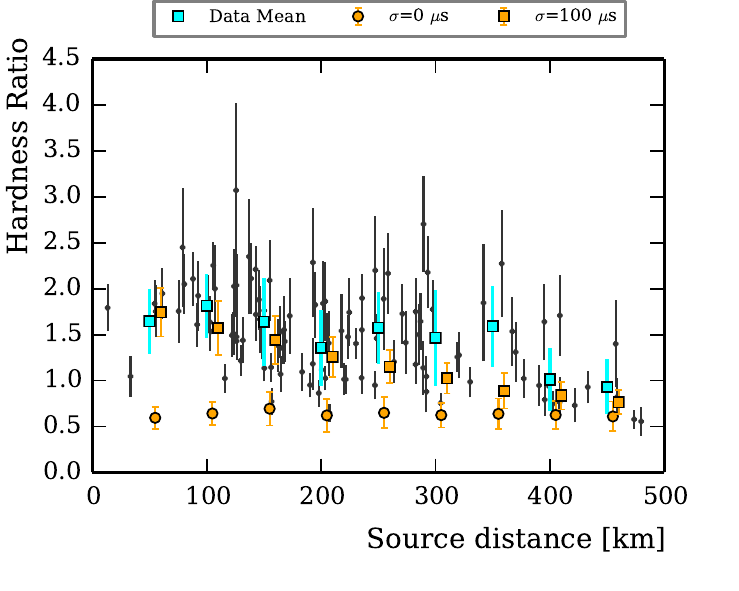}
	\caption{Hardness Ratio as a function of the distance between the source and the \textit{Fermi} sub-satellite position. The data are plotted in the same fashion as Figure~\ref{fig:delay}. The expected softening with increased offset is evident. The values derived from the simulations with no smearing are significantly softer than the data. Simulations with longer smearing  timescales ($>$50\,$\mu $s) are more representative of the data, but still tend to underestimate the hardness. In the interests of clarity, only the results from the 100\,$\mu $s simulations are overplotted.}
\label{fig:hr} \end{figure}

The hardness ratio as a function of the TGF source to sub-satellite location distance is shown in Figure~\ref{fig:hr} for the data and simulations. A clear trend of increased softening with offset is visible in the data. The simulations which underwent no smearing are completely inconsistent with the data. Simulations with a  25\,$\mu $s smearing factor are inconsistent with the data up to 150\,km. After this point they follow the trend but are significantly softer than the data mean. The simulations with longer smearings (50, 75 and 100\,$\mu $s) are consistent with each other, and agree best with the data, but are also systematically softer than the mean of the data. Possible explanations for this discrepancy are discussed in the following section.

\begin{figure}\center
	\noindent\includegraphics[width=8cm]{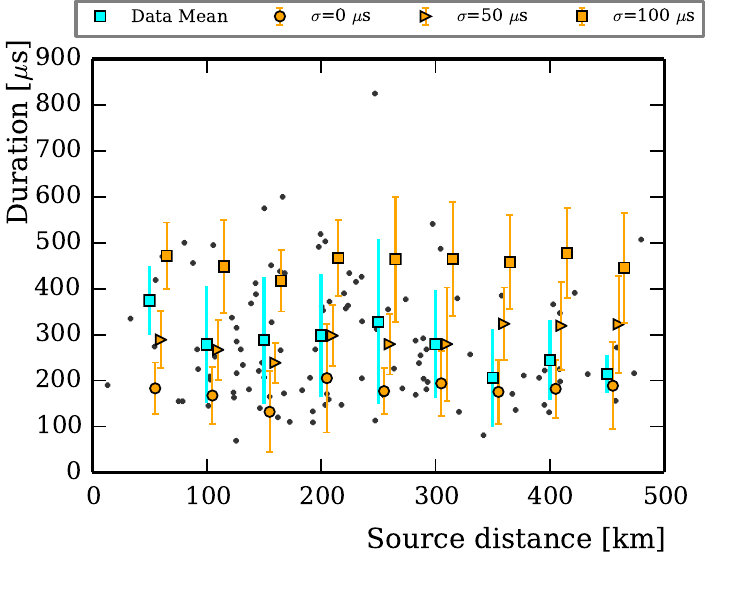}
	\caption{Calculated duration as a function of the distance between the source and the \textit{Fermi} sub-satellite position.  The data are plotted in the same fashion as Figure~\ref{fig:delay}. The durations from the simulations exhibit little variation with increased offset, and encompass the range of observed durations within their spread. The results from the 50\,$\mu $s smearing agree with the mean of the data up to 300\,km.}
\label{fig:duration} \end{figure}

The duration as a function of the TGF source to sub-satellite location distance is shown in Figure~\ref{fig:duration} for the data and simulations. The data exhibits a decreasing duration with increasing source distance, the opposite of the predicted relation. This is likely a consequence of the decrease in fluence with increasing source distance (Figure~\ref{f:flu}), as it will be more difficult to distinguish the TGF from the background, resulting in a shorter observed duration. The durations from the simulations exhibit little variation with increased offset, and encompass the range of observed durations within their spread. The results from the 50\,$\mu $s smearing agree with the mean of the data up to 300\,km, after which on they are consistently longer than the data.

\section{Discussion}
Assuming a wide beam model and a single altitude source, an increase in the off-axis distance is expected to cause a corresponding increase in duration and decrease in spectral hardness (see Figure~\ref{f:sim}). This temporal elongation is not observed in either the data or the simulations, which is likely to be due to the decrease in the incident fluence as the off-axis distance is increased. In contrast, the predicted softening with increased off-axis distance is evident in the data. The simulations with smearings less than 50\,$\mu $s are generally inconsistent with the mean of the data, implying that the source electron distribution for the majority of TGFs observed by GBM is not necessarily created instantaneously, and likely varies on timescales $\ge$\,50\,$\mu $s.

It has been hypothesised that short TGFs observed by GBM ($T_{50}<50$\,$\mu $s) are the result of instantaneous sources, and those with longer durations are the result of a superposition of these pulses \citep{Celestin2012}. This analysis would seem to challenge the former, as the spectrum obtained from a short electron pulse ($<25\mu$s) is inconsistent with the data. This implies that the temporal elongation due to Compton scattering is not sufficient, and that an intrinsic time variation of at least tens of $\mu$s is required at the source. 

Even for the largest smearing considered (100\,$\mu $s), the data exhibit harder spectra than can be explained by the simulations. We posit three possible explanations for this discrepancy in order of likelihood. The assumption of a single source altitude of 15\,km may be incorrect. If instead there are a range of source altitudes, then the evolution of HR with off-axis distance would be considerably broadened. TGFs with harder spectra may be simply explained by a higher source altitude, as the photons will undergo less Compton scattering due to the reduced integrated atmosphere traversed. Alternatively, the RREA model used in this work may be too basic, and more complex models which include an evolving electron source distribution (e.g. \citet{Dwyer2012a})  may be required to explain the range of observed values. Finally, `tilted' beams may be invoked. In this scenario, the alignment of the \textit{E} field at the source is not directly parallel to the vertical, with the result that the beam of gamma rays is correspondingly tilted. In such a case, the off-axis HR would be harder than expected due to the greater proportion of high energy photons. However, the same general trend of decreasing hardness with increased off-axis distance would remain, as the tilted beams would traverse a greater depth of atmosphere than the vertical equivalent. 

It also possible that the explanation is a combination of the first two points above. TGFs close to the satellite nadir could have a range of source altitudes with primarily vertical beams. As the source distance is increased, TGFs with higher altitude sources and/or tilted beams would be preferentially detected. 
Accounting for this possible selection effect would introduce degeneracies that cannot be mitigated with our sample size.  They will be investigated in future analyses using simulations and a larger sample size.

Our results imply that the majority of TGFs observed by GBM must have source electron distributions that vary on timescales of at least tens of \,$\mu $s. Based on Figure~\ref{fig:duration}, we suggest that 100\,$\mu $s is close to the upper limit of the source electron distribution variation timescale, with a value of 50\,$\mu $s being a likely mean. Of the five empirical smearing distributions considered, only the 50\,$\mu $s Gaussian is consistent with the observed temporal and spectral properties of the data (i.e. delay, duration and hardness). This consistency extends up to 200\,--\,250\,km, at which point the simulated times become longer and the spectra become softer. This could be attributed to insufficient statistics beyond this source-satellite distance (see Figure~\ref{f:flu}). 

In this work, we have performed a comprehensive study of RREA simulations in comparison to GBM observations. The observations exhibit a clear softening with increased source distance, in qualitative agreement with theoretical predictions. However, a quantitative analysis shows that the observed spectra can be harder than that predicted by the basic single source-altitude model. Simulations performed with an electron source timescale of 50\,\mus{} are most consistent with the temporal and spectral properties observed in the data. We propose that the source electron distributions of TGFs observed by GBM vary on timescales of at least tens of microseconds, with an upper limit of $\approx$100\,\mus{}. Drawing more concrete conclusions is limited by the low number of TGFs at larger off-axis source distances and the simplifying assumptions of the simulations. These assumptions, which are common in the literature, include a single fixed source altitude of 15 km and a vertically oriented E field. The effect of varying the source altitude and beam orientation will be investigated in a future work. 

\begin{acknowledgments}
The Fermi GBM Collaboration acknowledges the support of NASA in the United States and DRL in Germany. G.~F. acknowledges the support of the Irish Research Council. S.~MB. acknowledges support from Science Foundation Ireland under grant number 12/IP/1288.  The authors wish to thank the World Wide Lightning Location Network (http://wwlln.net), a collaboration among over 50 universities and institutions, for providing the lightning location data used in this paper. We also acknowledge Earth
Networks for providing the ENTLN data examined in this study.

\end{acknowledgments}
\bibliography{ref}

\begin{thebibliography}{33}%
\makeatletter
\providecommand \@ifxundefined [1]{%
 \@ifx{#1\undefined}
}%
\providecommand \@ifnum [1]{%
 \ifnum #1\expandafter \@firstoftwo
 \else \expandafter \@secondoftwo
 \fi
}%
\providecommand \@ifx [1]{%
 \ifx #1\expandafter \@firstoftwo
 \else \expandafter \@secondoftwo
 \fi
}%
\providecommand \natexlab [1]{#1}%
\providecommand \enquote  [1]{``#1''}%
\providecommand \bibnamefont  [1]{#1}%
\providecommand \bibfnamefont [1]{#1}%
\providecommand \citenamefont [1]{#1}%
\providecommand \href@noop [0]{\@secondoftwo}%
\providecommand \href [0]{\begingroup \@sanitize@url \@href}%
\providecommand \@href[1]{\@@startlink{#1}\@@href}%
\providecommand \@@href[1]{\endgroup#1\@@endlink}%
\providecommand \@sanitize@url [0]{\catcode `\\12\catcode `\$12\catcode
  `\&12\catcode `\#12\catcode `\^12\catcode `\_12\catcode `\%12\relax}%
\providecommand \@@startlink[1]{}%
\providecommand \@@endlink[0]{}%
\providecommand \url  [0]{\begingroup\@sanitize@url \@url }%
\providecommand \@url [1]{\endgroup\@href {#1}{\urlprefix }}%
\providecommand \urlprefix  [0]{URL }%
\providecommand \Eprint [0]{\href }%
\providecommand \doibase [0]{http://dx.doi.org/}%
\providecommand \selectlanguage [0]{\@gobble}%
\providecommand \bibinfo  [0]{\@secondoftwo}%
\providecommand \bibfield  [0]{\@secondoftwo}%
\providecommand \translation [1]{[#1]}%
\providecommand \BibitemOpen [0]{}%
\providecommand \bibitemStop [0]{}%
\providecommand \bibitemNoStop [0]{.\EOS\space}%
\providecommand \EOS [0]{\spacefactor3000\relax}%
\providecommand \BibitemShut  [1]{\csname bibitem#1\endcsname}%
\let\auto@bib@innerbib\@empty
\bibitem [{\citenamefont {Fishman}\ \emph {et~al.}(1994)\citenamefont
  {Fishman}, \citenamefont {Bhat}, \citenamefont {Mallozzi}, \citenamefont
  {Horack}, \citenamefont {Koshut}, \citenamefont {Kouveliotou}, \citenamefont
  {Pendleton}, \citenamefont {Meegan}, \citenamefont {Wilson}, \citenamefont
  {Paciesas}, \citenamefont {Goodman},\ and\ \citenamefont
  {Christian}}]{Fishman1994}%
  \BibitemOpen
  \bibfield  {author} {\bibinfo {author} {\bibfnamefont {G.~J.}\ \bibnamefont
  {Fishman}}, \bibinfo {author} {\bibfnamefont {P.~N.}\ \bibnamefont {Bhat}},
  \bibinfo {author} {\bibfnamefont {R.}~\bibnamefont {Mallozzi}}, \bibinfo
  {author} {\bibfnamefont {J.~M.}\ \bibnamefont {Horack}}, \bibinfo {author}
  {\bibfnamefont {T.}~\bibnamefont {Koshut}}, \bibinfo {author} {\bibfnamefont
  {C.}~\bibnamefont {Kouveliotou}}, \bibinfo {author} {\bibfnamefont {G.~N.}\
  \bibnamefont {Pendleton}}, \bibinfo {author} {\bibfnamefont {C.~a.}\
  \bibnamefont {Meegan}}, \bibinfo {author} {\bibfnamefont {R.~B.}\
  \bibnamefont {Wilson}}, \bibinfo {author} {\bibfnamefont {W.~S.}\
  \bibnamefont {Paciesas}}, \bibinfo {author} {\bibfnamefont {S.~J.}\
  \bibnamefont {Goodman}}, \ and\ \bibinfo {author} {\bibfnamefont {H.~J.}\
  \bibnamefont {Christian}},\ }\href {\doibase 10.1126/science.264.5163.1313}
  {\bibfield  {journal} {\bibinfo  {journal} {Science (New York, N.Y.)}\
  }\textbf {\bibinfo {volume} {264}},\ \bibinfo {pages} {1313} (\bibinfo {year}
  {1994})}\BibitemShut {NoStop}%
\bibitem [{\citenamefont {Smith}\ \emph {et~al.}(2005)\citenamefont {Smith},
  \citenamefont {Lopez}, \citenamefont {Lin},\ and\ \citenamefont
  {Barrington-Leigh}}]{Smith2005}%
  \BibitemOpen
  \bibfield  {author} {\bibinfo {author} {\bibfnamefont {D.~M.}\ \bibnamefont
  {Smith}}, \bibinfo {author} {\bibfnamefont {L.~I.}\ \bibnamefont {Lopez}},
  \bibinfo {author} {\bibfnamefont {R.~P.}\ \bibnamefont {Lin}}, \ and\
  \bibinfo {author} {\bibfnamefont {C.~P.}\ \bibnamefont {Barrington-Leigh}},\
  }\href {\doibase 10.1126/science.1107466} {\bibfield  {journal} {\bibinfo
  {journal} {Science (New York, N.Y.)}\ }\textbf {\bibinfo {volume} {307}},\
  \bibinfo {pages} {1085} (\bibinfo {year} {2005})}\BibitemShut {NoStop}%
\bibitem [{\citenamefont {Grefenstette}\ \emph {et~al.}(2009)\citenamefont
  {Grefenstette}, \citenamefont {Smith}, \citenamefont {Hazelton},\ and\
  \citenamefont {Lopez}}]{Grefenstette2009}%
  \BibitemOpen
  \bibfield  {author} {\bibinfo {author} {\bibfnamefont {B.~W.}\ \bibnamefont
  {Grefenstette}}, \bibinfo {author} {\bibfnamefont {D.~M.}\ \bibnamefont
  {Smith}}, \bibinfo {author} {\bibfnamefont {B.~J.}\ \bibnamefont {Hazelton}},
  \ and\ \bibinfo {author} {\bibfnamefont {L.~I.}\ \bibnamefont {Lopez}},\
  }\href {\doibase 10.1029/2008JA013721} {\bibfield  {journal} {\bibinfo
  {journal} {Journal of Geophysical Research}\ }\textbf {\bibinfo {volume}
  {114}},\ \bibinfo {pages} {A02314} (\bibinfo {year} {2009})}\BibitemShut
  {NoStop}%
\bibitem [{\citenamefont {Marisaldi}\ \emph {et~al.}(2013)\citenamefont
  {Marisaldi}, \citenamefont {Fuschino}, \citenamefont {Tavani}, \citenamefont
  {Dietrich}, \citenamefont {Price}, \citenamefont {Galli}, \citenamefont
  {Pittori}, \citenamefont {Verrecchia}, \citenamefont {Mereghetti},
  \citenamefont {Cattaneo}, \citenamefont {Colafrancesco}, \citenamefont
  {Argan}, \citenamefont {Labanti}, \citenamefont {Longo}, \citenamefont {{Del
  Monte}}, \citenamefont {Barbiellini}, \citenamefont {Giuliani}, \citenamefont
  {Bulgarelli}, \citenamefont {Campana}, \citenamefont {Chen}, \citenamefont
  {Gianotti}, \citenamefont {Giommi}, \citenamefont {Lazzarotto}, \citenamefont
  {Morselli}, \citenamefont {Rapisarda}, \citenamefont {Rappoldi},
  \citenamefont {Trifoglio}, \citenamefont {Trois},\ and\ \citenamefont
  {Vercellone}}]{Marisaldi2013}%
  \BibitemOpen
  \bibfield  {author} {\bibinfo {author} {\bibfnamefont {M.}~\bibnamefont
  {Marisaldi}}, \bibinfo {author} {\bibfnamefont {F.}~\bibnamefont {Fuschino}},
  \bibinfo {author} {\bibfnamefont {M.}~\bibnamefont {Tavani}}, \bibinfo
  {author} {\bibfnamefont {S.}~\bibnamefont {Dietrich}}, \bibinfo {author}
  {\bibfnamefont {C.}~\bibnamefont {Price}}, \bibinfo {author} {\bibfnamefont
  {M.}~\bibnamefont {Galli}}, \bibinfo {author} {\bibfnamefont
  {C.}~\bibnamefont {Pittori}}, \bibinfo {author} {\bibfnamefont
  {F.}~\bibnamefont {Verrecchia}}, \bibinfo {author} {\bibfnamefont
  {S.}~\bibnamefont {Mereghetti}}, \bibinfo {author} {\bibfnamefont
  {P.}~\bibnamefont {Cattaneo}}, \bibinfo {author} {\bibfnamefont
  {S.}~\bibnamefont {Colafrancesco}}, \bibinfo {author} {\bibfnamefont
  {A.}~\bibnamefont {Argan}}, \bibinfo {author} {\bibfnamefont
  {C.}~\bibnamefont {Labanti}}, \bibinfo {author} {\bibfnamefont
  {F.}~\bibnamefont {Longo}}, \bibinfo {author} {\bibfnamefont
  {E.}~\bibnamefont {{Del Monte}}}, \bibinfo {author} {\bibfnamefont
  {G.}~\bibnamefont {Barbiellini}}, \bibinfo {author} {\bibfnamefont
  {A.}~\bibnamefont {Giuliani}}, \bibinfo {author} {\bibfnamefont
  {A.}~\bibnamefont {Bulgarelli}}, \bibinfo {author} {\bibfnamefont
  {R.}~\bibnamefont {Campana}}, \bibinfo {author} {\bibfnamefont
  {A.}~\bibnamefont {Chen}}, \bibinfo {author} {\bibfnamefont {F.}~\bibnamefont
  {Gianotti}}, \bibinfo {author} {\bibfnamefont {P.}~\bibnamefont {Giommi}},
  \bibinfo {author} {\bibfnamefont {F.}~\bibnamefont {Lazzarotto}}, \bibinfo
  {author} {\bibfnamefont {A.}~\bibnamefont {Morselli}}, \bibinfo {author}
  {\bibfnamefont {M.}~\bibnamefont {Rapisarda}}, \bibinfo {author}
  {\bibfnamefont {A.}~\bibnamefont {Rappoldi}}, \bibinfo {author}
  {\bibfnamefont {M.}~\bibnamefont {Trifoglio}}, \bibinfo {author}
  {\bibfnamefont {A.}~\bibnamefont {Trois}}, \ and\ \bibinfo {author}
  {\bibfnamefont {S.}~\bibnamefont {Vercellone}},\ }\href {\doibase
  10.1002/2013JA019301} {\bibfield  {journal} {\bibinfo  {journal} {Journal of
  Geophysical Research: Space Physics}\ } (\bibinfo {year} {2013}),\
  10.1002/2013JA019301}\BibitemShut {NoStop}%
\bibitem [{\citenamefont {Briggs}(2013)}]{Briggs2013}%
  \BibitemOpen
  \bibfield  {author} {\bibinfo {author} {\bibfnamefont {M.}~\bibnamefont
  {Briggs}},\ }\href {\doibase 10.1002/jgra.50205} {\bibfield  {journal}
  {\bibinfo  {journal} {Journal of Geophysical Research: Space Physics}\ }
  (\bibinfo {year} {2013}),\ 10.1002/jgra.50205}\BibitemShut {NoStop}%
\bibitem [{\citenamefont {{Schaal}}\ \emph {et~al.}(2013)\citenamefont
  {{Schaal}}, \citenamefont {{Grove}}, \citenamefont {{Chekhtman}},
  \citenamefont {{Xiong}}, \citenamefont {{Fitzpatrick}}, \citenamefont
  {{Cummer}},\ and\ \citenamefont {{Holzworth}}}]{GROVE_REF}%
  \BibitemOpen
  \bibfield  {author} {\bibinfo {author} {\bibfnamefont {M.}~\bibnamefont
  {{Schaal}}}, \bibinfo {author} {\bibfnamefont {J.}~\bibnamefont {{Grove}}},
  \bibinfo {author} {\bibfnamefont {A.}~\bibnamefont {{Chekhtman}}}, \bibinfo
  {author} {\bibfnamefont {S.}~\bibnamefont {{Xiong}}}, \bibinfo {author}
  {\bibfnamefont {G.}~\bibnamefont {{Fitzpatrick}}}, \bibinfo {author}
  {\bibfnamefont {S.~A.}\ \bibnamefont {{Cummer}}}, \ and\ \bibinfo {author}
  {\bibfnamefont {R.~H.}\ \bibnamefont {{Holzworth}}},\ }\href@noop {}
  {\bibfield  {journal} {\bibinfo  {journal} {AGU Fall Meeting Abstracts}\ ,\
  \bibinfo {pages} {A405}} (\bibinfo {year} {2013})}\BibitemShut {NoStop}%
\bibitem [{\citenamefont {Gurevich}\ \emph {et~al.}(1992)\citenamefont
  {Gurevich}, \citenamefont {Milikh},\ and\ \citenamefont
  {Roussel-Dupre}}]{1992PhLA..165..463G}%
  \BibitemOpen
  \bibfield  {author} {\bibinfo {author} {\bibfnamefont {A.~V.}\ \bibnamefont
  {Gurevich}}, \bibinfo {author} {\bibfnamefont {G.~M.}\ \bibnamefont
  {Milikh}}, \ and\ \bibinfo {author} {\bibfnamefont {R.}~\bibnamefont
  {Roussel-Dupre}},\ }\href {\doibase 10.1016/0375-9601(92)90348-P} {\bibfield
  {journal} {\bibinfo  {journal} {Physics Letters A}\ }\textbf {\bibinfo
  {volume} {165}},\ \bibinfo {pages} {463} (\bibinfo {year}
  {1992})}\BibitemShut {NoStop}%
\bibitem [{\citenamefont {Grefenstette}\ \emph {et~al.}(2008)\citenamefont
  {Grefenstette}, \citenamefont {Smith}, \citenamefont {Dwyer},\ and\
  \citenamefont {Fishman}}]{2008GeoRL..3506802G}%
  \BibitemOpen
  \bibfield  {author} {\bibinfo {author} {\bibfnamefont {B.~W.}\ \bibnamefont
  {Grefenstette}}, \bibinfo {author} {\bibfnamefont {D.~M.}\ \bibnamefont
  {Smith}}, \bibinfo {author} {\bibfnamefont {J.~R.}\ \bibnamefont {Dwyer}}, \
  and\ \bibinfo {author} {\bibfnamefont {G.~J.}\ \bibnamefont {Fishman}},\
  }\href {\doibase 10.1029/2007GL032922} {\bibfield  {journal} {\bibinfo
  {journal} {Geophysical Research Letters}\ }\textbf {\bibinfo {volume} {35}},\
  \bibinfo {pages} {L06802} (\bibinfo {year} {2008})}\BibitemShut {NoStop}%
\bibitem [{\citenamefont {Marisaldi}\ \emph {et~al.}(2010)\citenamefont
  {Marisaldi}, \citenamefont {Fuschino}, \citenamefont {Labanti}, \citenamefont
  {Galli}, \citenamefont {Longo}, \citenamefont {{Del Monte}}, \citenamefont
  {Barbiellini}, \citenamefont {Tavani}, \citenamefont {Giuliani},
  \citenamefont {Moretti}, \citenamefont {Vercellone}, \citenamefont {Costa},
  \citenamefont {Cutini}, \citenamefont {Donnarumma}, \citenamefont
  {Evangelista}, \citenamefont {Feroci}, \citenamefont {Lapshov}, \citenamefont
  {Lazzarotto}, \citenamefont {Lipari}, \citenamefont {Mereghetti},
  \citenamefont {Pacciani}, \citenamefont {Rapisarda}, \citenamefont
  {Soffitta}, \citenamefont {Trifoglio}, \citenamefont {Argan}, \citenamefont
  {Boffelli}, \citenamefont {Bulgarelli}, \citenamefont {Caraveo},
  \citenamefont {Cattaneo}, \citenamefont {Chen}, \citenamefont {Cocco},
  \citenamefont {D'Ammando}, \citenamefont {{De Paris}}, \citenamefont {{Di
  Cocco}}, \citenamefont {{Di Persio}}, \citenamefont {Ferrari}, \citenamefont
  {Fiorini}, \citenamefont {Froysland}, \citenamefont {Gianotti}, \citenamefont
  {Morselli}, \citenamefont {Pellizzoni}, \citenamefont {Perotti},
  \citenamefont {Picozza}, \citenamefont {Piano}, \citenamefont {Pilia},
  \citenamefont {Prest}, \citenamefont {Pucella}, \citenamefont {Rappoldi},
  \citenamefont {Rubini}, \citenamefont {Sabatini}, \citenamefont {Striani},
  \citenamefont {Trois}, \citenamefont {Vallazza}, \citenamefont {Vittorini},
  \citenamefont {Zambra}, \citenamefont {Zanello}, \citenamefont {Antonelli},
  \citenamefont {Colafrancesco}, \citenamefont {Gasparrini}, \citenamefont
  {Giommi}, \citenamefont {Pittori}, \citenamefont {Preger}, \citenamefont
  {Santolamazza}, \citenamefont {Verrecchia},\ and\ \citenamefont
  {Salotti}}]{Marisaldi2010a}%
  \BibitemOpen
  \bibfield  {author} {\bibinfo {author} {\bibfnamefont {M.}~\bibnamefont
  {Marisaldi}}, \bibinfo {author} {\bibfnamefont {F.}~\bibnamefont {Fuschino}},
  \bibinfo {author} {\bibfnamefont {C.}~\bibnamefont {Labanti}}, \bibinfo
  {author} {\bibfnamefont {M.}~\bibnamefont {Galli}}, \bibinfo {author}
  {\bibfnamefont {F.}~\bibnamefont {Longo}}, \bibinfo {author} {\bibfnamefont
  {E.}~\bibnamefont {{Del Monte}}}, \bibinfo {author} {\bibfnamefont
  {G.}~\bibnamefont {Barbiellini}}, \bibinfo {author} {\bibfnamefont
  {M.}~\bibnamefont {Tavani}}, \bibinfo {author} {\bibfnamefont
  {A.}~\bibnamefont {Giuliani}}, \bibinfo {author} {\bibfnamefont
  {E.}~\bibnamefont {Moretti}}, \bibinfo {author} {\bibfnamefont
  {S.}~\bibnamefont {Vercellone}}, \bibinfo {author} {\bibfnamefont
  {E.}~\bibnamefont {Costa}}, \bibinfo {author} {\bibfnamefont
  {S.}~\bibnamefont {Cutini}}, \bibinfo {author} {\bibfnamefont
  {I.}~\bibnamefont {Donnarumma}}, \bibinfo {author} {\bibfnamefont
  {Y.}~\bibnamefont {Evangelista}}, \bibinfo {author} {\bibfnamefont
  {M.}~\bibnamefont {Feroci}}, \bibinfo {author} {\bibfnamefont
  {I.}~\bibnamefont {Lapshov}}, \bibinfo {author} {\bibfnamefont
  {F.}~\bibnamefont {Lazzarotto}}, \bibinfo {author} {\bibfnamefont
  {P.}~\bibnamefont {Lipari}}, \bibinfo {author} {\bibfnamefont
  {S.}~\bibnamefont {Mereghetti}}, \bibinfo {author} {\bibfnamefont
  {L.}~\bibnamefont {Pacciani}}, \bibinfo {author} {\bibfnamefont
  {M.}~\bibnamefont {Rapisarda}}, \bibinfo {author} {\bibfnamefont
  {P.}~\bibnamefont {Soffitta}}, \bibinfo {author} {\bibfnamefont
  {M.}~\bibnamefont {Trifoglio}}, \bibinfo {author} {\bibfnamefont
  {A.}~\bibnamefont {Argan}}, \bibinfo {author} {\bibfnamefont
  {F.}~\bibnamefont {Boffelli}}, \bibinfo {author} {\bibfnamefont
  {A.}~\bibnamefont {Bulgarelli}}, \bibinfo {author} {\bibfnamefont
  {P.}~\bibnamefont {Caraveo}}, \bibinfo {author} {\bibfnamefont {P.~W.}\
  \bibnamefont {Cattaneo}}, \bibinfo {author} {\bibfnamefont {A.}~\bibnamefont
  {Chen}}, \bibinfo {author} {\bibfnamefont {V.}~\bibnamefont {Cocco}},
  \bibinfo {author} {\bibfnamefont {F.}~\bibnamefont {D'Ammando}}, \bibinfo
  {author} {\bibfnamefont {G.}~\bibnamefont {{De Paris}}}, \bibinfo {author}
  {\bibfnamefont {G.}~\bibnamefont {{Di Cocco}}}, \bibinfo {author}
  {\bibfnamefont {G.}~\bibnamefont {{Di Persio}}}, \bibinfo {author}
  {\bibfnamefont {A.}~\bibnamefont {Ferrari}}, \bibinfo {author} {\bibfnamefont
  {M.}~\bibnamefont {Fiorini}}, \bibinfo {author} {\bibfnamefont
  {T.}~\bibnamefont {Froysland}}, \bibinfo {author} {\bibfnamefont
  {F.}~\bibnamefont {Gianotti}}, \bibinfo {author} {\bibfnamefont
  {A.}~\bibnamefont {Morselli}}, \bibinfo {author} {\bibfnamefont
  {A.}~\bibnamefont {Pellizzoni}}, \bibinfo {author} {\bibfnamefont
  {F.}~\bibnamefont {Perotti}}, \bibinfo {author} {\bibfnamefont
  {P.}~\bibnamefont {Picozza}}, \bibinfo {author} {\bibfnamefont
  {G.}~\bibnamefont {Piano}}, \bibinfo {author} {\bibfnamefont
  {M.}~\bibnamefont {Pilia}}, \bibinfo {author} {\bibfnamefont
  {M.}~\bibnamefont {Prest}}, \bibinfo {author} {\bibfnamefont
  {G.}~\bibnamefont {Pucella}}, \bibinfo {author} {\bibfnamefont
  {A.}~\bibnamefont {Rappoldi}}, \bibinfo {author} {\bibfnamefont
  {A.}~\bibnamefont {Rubini}}, \bibinfo {author} {\bibfnamefont
  {S.}~\bibnamefont {Sabatini}}, \bibinfo {author} {\bibfnamefont
  {E.}~\bibnamefont {Striani}}, \bibinfo {author} {\bibfnamefont
  {A.}~\bibnamefont {Trois}}, \bibinfo {author} {\bibfnamefont
  {E.}~\bibnamefont {Vallazza}}, \bibinfo {author} {\bibfnamefont
  {V.}~\bibnamefont {Vittorini}}, \bibinfo {author} {\bibfnamefont
  {A.}~\bibnamefont {Zambra}}, \bibinfo {author} {\bibfnamefont
  {D.}~\bibnamefont {Zanello}}, \bibinfo {author} {\bibfnamefont {L.~a.}\
  \bibnamefont {Antonelli}}, \bibinfo {author} {\bibfnamefont {S.}~\bibnamefont
  {Colafrancesco}}, \bibinfo {author} {\bibfnamefont {D.}~\bibnamefont
  {Gasparrini}}, \bibinfo {author} {\bibfnamefont {P.}~\bibnamefont {Giommi}},
  \bibinfo {author} {\bibfnamefont {C.}~\bibnamefont {Pittori}}, \bibinfo
  {author} {\bibfnamefont {B.}~\bibnamefont {Preger}}, \bibinfo {author}
  {\bibfnamefont {P.}~\bibnamefont {Santolamazza}}, \bibinfo {author}
  {\bibfnamefont {F.}~\bibnamefont {Verrecchia}}, \ and\ \bibinfo {author}
  {\bibfnamefont {L.}~\bibnamefont {Salotti}},\ }\href {\doibase
  10.1029/2009JA014502} {\bibfield  {journal} {\bibinfo  {journal} {Journal of
  Geophysical Research}\ }\textbf {\bibinfo {volume} {115}},\ \bibinfo {pages}
  {1} (\bibinfo {year} {2010})}\BibitemShut {NoStop}%
\bibitem [{\citenamefont {Tavani}\ \emph {et~al.}(2011)\citenamefont {Tavani},
  \citenamefont {Marisaldi}, \citenamefont {Labanti}, \citenamefont {Fuschino},
  \citenamefont {Argan}, \citenamefont {Trois}, \citenamefont {Giommi},
  \citenamefont {Colafrancesco}, \citenamefont {Pittori}, \citenamefont
  {Palma}, \citenamefont {Trifoglio}, \citenamefont {Gianotti}, \citenamefont
  {Bulgarelli}, \citenamefont {Vittorini}, \citenamefont {Verrecchia},
  \citenamefont {Salotti}, \citenamefont {Barbiellini}, \citenamefont
  {Caraveo}, \citenamefont {Cattaneo}, \citenamefont {Chen}, \citenamefont
  {Contessi}, \citenamefont {Costa}, \citenamefont {D’Ammando}, \citenamefont
  {{Del Monte}}, \citenamefont {{De Paris}}, \citenamefont {{Di Cocco}},
  \citenamefont {{Di Persio}}, \citenamefont {Donnarumma}, \citenamefont
  {Evangelista}, \citenamefont {Feroci}, \citenamefont {Ferrari}, \citenamefont
  {Galli}, \citenamefont {Giuliani}, \citenamefont {Giusti}, \citenamefont
  {Lapshov}, \citenamefont {Lazzarotto}, \citenamefont {Lipari}, \citenamefont
  {Longo}, \citenamefont {Mereghetti}, \citenamefont {Morelli}, \citenamefont
  {Moretti}, \citenamefont {Morselli}, \citenamefont {Pacciani}, \citenamefont
  {Pellizzoni}, \citenamefont {Perotti}, \citenamefont {Piano}, \citenamefont
  {Picozza}, \citenamefont {Pilia}, \citenamefont {Pucella}, \citenamefont
  {Prest}, \citenamefont {Rapisarda}, \citenamefont {Rappoldi}, \citenamefont
  {Rossi}, \citenamefont {Rubini}, \citenamefont {Sabatini}, \citenamefont
  {Scalise}, \citenamefont {Soffitta}, \citenamefont {Striani}, \citenamefont
  {Vallazza}, \citenamefont {Vercellone}, \citenamefont {Zambra},\ and\
  \citenamefont {Zanello}}]{Tavani2011}%
  \BibitemOpen
  \bibfield  {author} {\bibinfo {author} {\bibfnamefont {M.}~\bibnamefont
  {Tavani}}, \bibinfo {author} {\bibfnamefont {M.}~\bibnamefont {Marisaldi}},
  \bibinfo {author} {\bibfnamefont {C.}~\bibnamefont {Labanti}}, \bibinfo
  {author} {\bibfnamefont {F.}~\bibnamefont {Fuschino}}, \bibinfo {author}
  {\bibfnamefont {A.}~\bibnamefont {Argan}}, \bibinfo {author} {\bibfnamefont
  {A.}~\bibnamefont {Trois}}, \bibinfo {author} {\bibfnamefont
  {P.}~\bibnamefont {Giommi}}, \bibinfo {author} {\bibfnamefont
  {S.}~\bibnamefont {Colafrancesco}}, \bibinfo {author} {\bibfnamefont
  {C.}~\bibnamefont {Pittori}}, \bibinfo {author} {\bibfnamefont
  {F.}~\bibnamefont {Palma}}, \bibinfo {author} {\bibfnamefont
  {M.}~\bibnamefont {Trifoglio}}, \bibinfo {author} {\bibfnamefont
  {F.}~\bibnamefont {Gianotti}}, \bibinfo {author} {\bibfnamefont
  {A.}~\bibnamefont {Bulgarelli}}, \bibinfo {author} {\bibfnamefont
  {V.}~\bibnamefont {Vittorini}}, \bibinfo {author} {\bibfnamefont
  {F.}~\bibnamefont {Verrecchia}}, \bibinfo {author} {\bibfnamefont
  {L.}~\bibnamefont {Salotti}}, \bibinfo {author} {\bibfnamefont
  {G.}~\bibnamefont {Barbiellini}}, \bibinfo {author} {\bibfnamefont
  {P.}~\bibnamefont {Caraveo}}, \bibinfo {author} {\bibfnamefont {P.~W.}\
  \bibnamefont {Cattaneo}}, \bibinfo {author} {\bibfnamefont {A.}~\bibnamefont
  {Chen}}, \bibinfo {author} {\bibfnamefont {T.}~\bibnamefont {Contessi}},
  \bibinfo {author} {\bibfnamefont {E.}~\bibnamefont {Costa}}, \bibinfo
  {author} {\bibfnamefont {F.}~\bibnamefont {D’Ammando}}, \bibinfo {author}
  {\bibfnamefont {E.}~\bibnamefont {{Del Monte}}}, \bibinfo {author}
  {\bibfnamefont {G.}~\bibnamefont {{De Paris}}}, \bibinfo {author}
  {\bibfnamefont {G.}~\bibnamefont {{Di Cocco}}}, \bibinfo {author}
  {\bibfnamefont {G.}~\bibnamefont {{Di Persio}}}, \bibinfo {author}
  {\bibfnamefont {I.}~\bibnamefont {Donnarumma}}, \bibinfo {author}
  {\bibfnamefont {Y.}~\bibnamefont {Evangelista}}, \bibinfo {author}
  {\bibfnamefont {M.}~\bibnamefont {Feroci}}, \bibinfo {author} {\bibfnamefont
  {A.}~\bibnamefont {Ferrari}}, \bibinfo {author} {\bibfnamefont
  {M.}~\bibnamefont {Galli}}, \bibinfo {author} {\bibfnamefont
  {A.}~\bibnamefont {Giuliani}}, \bibinfo {author} {\bibfnamefont
  {M.}~\bibnamefont {Giusti}}, \bibinfo {author} {\bibfnamefont
  {I.}~\bibnamefont {Lapshov}}, \bibinfo {author} {\bibfnamefont
  {F.}~\bibnamefont {Lazzarotto}}, \bibinfo {author} {\bibfnamefont
  {P.}~\bibnamefont {Lipari}}, \bibinfo {author} {\bibfnamefont
  {F.}~\bibnamefont {Longo}}, \bibinfo {author} {\bibfnamefont
  {S.}~\bibnamefont {Mereghetti}}, \bibinfo {author} {\bibfnamefont
  {E.}~\bibnamefont {Morelli}}, \bibinfo {author} {\bibfnamefont
  {E.}~\bibnamefont {Moretti}}, \bibinfo {author} {\bibfnamefont
  {A.}~\bibnamefont {Morselli}}, \bibinfo {author} {\bibfnamefont
  {L.}~\bibnamefont {Pacciani}}, \bibinfo {author} {\bibfnamefont
  {A.}~\bibnamefont {Pellizzoni}}, \bibinfo {author} {\bibfnamefont
  {F.}~\bibnamefont {Perotti}}, \bibinfo {author} {\bibfnamefont
  {G.}~\bibnamefont {Piano}}, \bibinfo {author} {\bibfnamefont
  {P.}~\bibnamefont {Picozza}}, \bibinfo {author} {\bibfnamefont
  {M.}~\bibnamefont {Pilia}}, \bibinfo {author} {\bibfnamefont
  {G.}~\bibnamefont {Pucella}}, \bibinfo {author} {\bibfnamefont
  {M.}~\bibnamefont {Prest}}, \bibinfo {author} {\bibfnamefont
  {M.}~\bibnamefont {Rapisarda}}, \bibinfo {author} {\bibfnamefont
  {A.}~\bibnamefont {Rappoldi}}, \bibinfo {author} {\bibfnamefont
  {E.}~\bibnamefont {Rossi}}, \bibinfo {author} {\bibfnamefont
  {A.}~\bibnamefont {Rubini}}, \bibinfo {author} {\bibfnamefont
  {S.}~\bibnamefont {Sabatini}}, \bibinfo {author} {\bibfnamefont
  {E.}~\bibnamefont {Scalise}}, \bibinfo {author} {\bibfnamefont
  {P.}~\bibnamefont {Soffitta}}, \bibinfo {author} {\bibfnamefont
  {E.}~\bibnamefont {Striani}}, \bibinfo {author} {\bibfnamefont
  {E.}~\bibnamefont {Vallazza}}, \bibinfo {author} {\bibfnamefont
  {S.}~\bibnamefont {Vercellone}}, \bibinfo {author} {\bibfnamefont
  {A.}~\bibnamefont {Zambra}}, \ and\ \bibinfo {author} {\bibfnamefont
  {D.}~\bibnamefont {Zanello}},\ }\href {\doibase
  10.1103/PhysRevLett.106.018501} {\bibfield  {journal} {\bibinfo  {journal}
  {Physical Review Letters}\ }\textbf {\bibinfo {volume} {106}},\ \bibinfo
  {pages} {018501} (\bibinfo {year} {2011})}\BibitemShut {NoStop}%
\bibitem [{\citenamefont {Feng}\ \emph {et~al.}(2002)\citenamefont {Feng},
  \citenamefont {Li}, \citenamefont {Wu}, \citenamefont {Zha},\ and\
  \citenamefont {Zhu}}]{Feng2002}%
  \BibitemOpen
  \bibfield  {author} {\bibinfo {author} {\bibfnamefont {H.}~\bibnamefont
  {Feng}}, \bibinfo {author} {\bibfnamefont {T.~P.}\ \bibnamefont {Li}},
  \bibinfo {author} {\bibfnamefont {M.}~\bibnamefont {Wu}}, \bibinfo {author}
  {\bibfnamefont {M.}~\bibnamefont {Zha}}, \ and\ \bibinfo {author}
  {\bibfnamefont {Q.~Q.}\ \bibnamefont {Zhu}},\ }\href {\doibase
  10.1029/2001GL013992} {\bibfield  {journal} {\bibinfo  {journal} {Geophysical
  Research Letters}\ }\textbf {\bibinfo {volume} {29}},\ \bibinfo {pages}
  {1036} (\bibinfo {year} {2002})}\BibitemShut {NoStop}%
\bibitem [{\citenamefont {Nemiroff}\ \emph {et~al.}(1997)\citenamefont
  {Nemiroff}, \citenamefont {Bonnell},\ and\ \citenamefont
  {Norris}}]{Nemiroff1997}%
  \BibitemOpen
  \bibfield  {author} {\bibinfo {author} {\bibfnamefont {R.~J.}\ \bibnamefont
  {Nemiroff}}, \bibinfo {author} {\bibfnamefont {J.~T.}\ \bibnamefont
  {Bonnell}}, \ and\ \bibinfo {author} {\bibfnamefont {J.~P.}\ \bibnamefont
  {Norris}},\ }\href {\doibase 10.1029/96JA03107} {\bibfield  {journal}
  {\bibinfo  {journal} {Journal of Geophysical Research}\ }\textbf {\bibinfo
  {volume} {102}},\ \bibinfo {pages} {9659} (\bibinfo {year}
  {1997})}\BibitemShut {NoStop}%
\bibitem [{\citenamefont {{\O stgaard}}\ \emph {et~al.}(2008)\citenamefont {{\O
  stgaard}}, \citenamefont {Gjesteland}, \citenamefont {Stadsnes},
  \citenamefont {Connell},\ and\ \citenamefont {Carlson}}]{Ostgaard2008}%
  \BibitemOpen
  \bibfield  {author} {\bibinfo {author} {\bibfnamefont {N.}~\bibnamefont {{\O
  stgaard}}}, \bibinfo {author} {\bibfnamefont {T.}~\bibnamefont {Gjesteland}},
  \bibinfo {author} {\bibfnamefont {J.}~\bibnamefont {Stadsnes}}, \bibinfo
  {author} {\bibfnamefont {P.~H.}\ \bibnamefont {Connell}}, \ and\ \bibinfo
  {author} {\bibfnamefont {B.}~\bibnamefont {Carlson}},\ }\href {\doibase
  10.1029/2007JA012618} {\bibfield  {journal} {\bibinfo  {journal} {Journal of
  Geophysical Research}\ }\textbf {\bibinfo {volume} {113}},\ \bibinfo {pages}
  {A02307} (\bibinfo {year} {2008})}\BibitemShut {NoStop}%
\bibitem [{\citenamefont {Gjesteland}\ \emph {et~al.}(2010)\citenamefont
  {Gjesteland}, \citenamefont {\O{}stgaard}, \citenamefont {Connell},
  \citenamefont {Stadsnes},\ and\ \citenamefont {Fishman}}]{Gjesteland2010}%
  \BibitemOpen
  \bibfield  {author} {\bibinfo {author} {\bibfnamefont {T.}~\bibnamefont
  {Gjesteland}}, \bibinfo {author} {\bibfnamefont {N.}~\bibnamefont
  {\O{}stgaard}}, \bibinfo {author} {\bibfnamefont {P.~H.}\ \bibnamefont
  {Connell}}, \bibinfo {author} {\bibfnamefont {J.}~\bibnamefont {Stadsnes}}, \
  and\ \bibinfo {author} {\bibfnamefont {G.~J.}\ \bibnamefont {Fishman}},\
  }\href {\doibase 10.1029/2009JA014578} {\bibfield  {journal} {\bibinfo
  {journal} {Journal of Geophysical Research}\ }\textbf {\bibinfo {volume}
  {115}},\ \bibinfo {pages} {A00E21} (\bibinfo {year} {2010})}\BibitemShut
  {NoStop}%
\bibitem [{\citenamefont {Celestin}\ and\ \citenamefont
  {Pasko}(2011)}]{2011JGRA..11603315C}%
  \BibitemOpen
  \bibfield  {author} {\bibinfo {author} {\bibfnamefont {S.}~\bibnamefont
  {Celestin}}\ and\ \bibinfo {author} {\bibfnamefont {V.~P.}\ \bibnamefont
  {Pasko}},\ }\href {\doibase 10.1029/2010JA016260} {\bibfield  {journal}
  {\bibinfo  {journal} {Journal of Geophysical Research}\ }\textbf {\bibinfo
  {volume} {116}},\ \bibinfo {pages} {A03315} (\bibinfo {year}
  {2011})}\BibitemShut {NoStop}%
\bibitem [{\citenamefont {Celestin}\ and\ \citenamefont
  {Pasko}(2012)}]{Celestin2012}%
  \BibitemOpen
  \bibfield  {author} {\bibinfo {author} {\bibfnamefont {S.}~\bibnamefont
  {Celestin}}\ and\ \bibinfo {author} {\bibfnamefont {V.~P.}\ \bibnamefont
  {Pasko}},\ }\href {\doibase 10.1029/2011GL050342} {\bibfield  {journal}
  {\bibinfo  {journal} {Geophysical Research Letters}\ }\textbf {\bibinfo
  {volume} {39}},\ \bibinfo {pages} {L02802} (\bibinfo {year}
  {2012})}\BibitemShut {NoStop}%
\bibitem [{\citenamefont {Foley}(ress)}]{Foley2014}%
  \BibitemOpen
  \bibfield  {author} {\bibinfo {author} {\bibfnamefont {S.}~\bibnamefont
  {Foley}},\ }\href@noop {} {\bibfield  {journal} {\bibinfo  {journal} {Journal
  of Geophysical Research: Space Physics}\ } (\bibinfo {year} {In
  press})}\BibitemShut {NoStop}%
\bibitem [{\citenamefont {Connaughton}\ \emph {et~al.}(2012)\citenamefont
  {Connaughton}, \citenamefont {Foley}, \citenamefont {McBreen}, \citenamefont
  {Bhat}, \citenamefont {Chaplin}, \citenamefont {Cramer}, \citenamefont
  {Fishman}, \citenamefont {Holzworth}, \citenamefont {Gibby}, \citenamefont
  {von Kienlin}, \citenamefont {Meegan}, \citenamefont {Briggs}, \citenamefont
  {Paciesas}, \citenamefont {Preece}, \citenamefont {Wilson-Hodge},
  \citenamefont {Xiong}, \citenamefont {Dwyer}, \citenamefont {Hutchins},
  \citenamefont {Grove}, \citenamefont {Chekhtman}, \citenamefont {Tierney},\
  and\ \citenamefont {Fitzpatrick}}]{Connaughton2012}%
  \BibitemOpen
  \bibfield  {author} {\bibinfo {author} {\bibfnamefont {V.}~\bibnamefont
  {Connaughton}}, \bibinfo {author} {\bibfnamefont {S.}~\bibnamefont {Foley}},
  \bibinfo {author} {\bibfnamefont {S.}~\bibnamefont {McBreen}}, \bibinfo
  {author} {\bibfnamefont {N.}~\bibnamefont {Bhat}}, \bibinfo {author}
  {\bibfnamefont {V.}~\bibnamefont {Chaplin}}, \bibinfo {author} {\bibfnamefont
  {E.}~\bibnamefont {Cramer}}, \bibinfo {author} {\bibfnamefont {G.~J.}\
  \bibnamefont {Fishman}}, \bibinfo {author} {\bibfnamefont {R.~H.}\
  \bibnamefont {Holzworth}}, \bibinfo {author} {\bibfnamefont {M.}~\bibnamefont
  {Gibby}}, \bibinfo {author} {\bibfnamefont {A.}~\bibnamefont {von Kienlin}},
  \bibinfo {author} {\bibfnamefont {C.~A.}\ \bibnamefont {Meegan}}, \bibinfo
  {author} {\bibfnamefont {M.~S.}\ \bibnamefont {Briggs}}, \bibinfo {author}
  {\bibfnamefont {W.~S.}\ \bibnamefont {Paciesas}}, \bibinfo {author}
  {\bibfnamefont {R.}~\bibnamefont {Preece}}, \bibinfo {author} {\bibfnamefont
  {C.}~\bibnamefont {Wilson-Hodge}}, \bibinfo {author} {\bibfnamefont
  {S.}~\bibnamefont {Xiong}}, \bibinfo {author} {\bibfnamefont {J.~R.}\
  \bibnamefont {Dwyer}}, \bibinfo {author} {\bibfnamefont {M.~L.}\ \bibnamefont
  {Hutchins}}, \bibinfo {author} {\bibfnamefont {J.~E.}\ \bibnamefont {Grove}},
  \bibinfo {author} {\bibfnamefont {A.}~\bibnamefont {Chekhtman}}, \bibinfo
  {author} {\bibfnamefont {D.}~\bibnamefont {Tierney}}, \ and\ \bibinfo
  {author} {\bibfnamefont {G.}~\bibnamefont {Fitzpatrick}},\ }\href {\doibase
  10.1029/2012JA018288} {\bibfield  {journal} {\bibinfo  {journal} {Journal of
  Geophysical Research}\ } (\bibinfo {year} {2012}),\
  10.1029/2012JA018288}\BibitemShut {NoStop}%
\bibitem [{\citenamefont {Dwyer}(2003)}]{Dwyer2003}%
  \BibitemOpen
  \bibfield  {author} {\bibinfo {author} {\bibfnamefont {J.~R.}\ \bibnamefont
  {Dwyer}},\ }\href {\doibase 10.1029/2003GL017781} {\bibfield  {journal}
  {\bibinfo  {journal} {Geophysical Research Letters}\ }\textbf {\bibinfo
  {volume} {30}},\ \bibinfo {pages} {1} (\bibinfo {year} {2003})}\BibitemShut
  {NoStop}%
\bibitem [{\citenamefont {Dwyer}(2007)}]{Dwyer2007}%
  \BibitemOpen
  \bibfield  {author} {\bibinfo {author} {\bibfnamefont {J.~R.}\ \bibnamefont
  {Dwyer}},\ }\href {\doibase 10.1063/1.2709652} {\bibfield  {journal}
  {\bibinfo  {journal} {Physics of Plasmas}\ }\textbf {\bibinfo {volume}
  {14}},\ \bibinfo {pages} {042901} (\bibinfo {year} {2007})}\BibitemShut
  {NoStop}%
\bibitem [{\citenamefont {Atwood}\ \emph {et~al.}(2009)\citenamefont {Atwood},
  \citenamefont {Abdo}, \citenamefont {Ackermann}, \citenamefont {Althouse},
  \citenamefont {Anderson}, \citenamefont {Axelsson}, \citenamefont {Baldini},
  \citenamefont {Ballet}, \citenamefont {Band}, \citenamefont {Barbiellini},
  \citenamefont {Bartelt}, \citenamefont {Bastieri}, \citenamefont {Baughman},
  \citenamefont {Bechtol}, \citenamefont {B\'{e}d\'{e}r\`{e}de}, \citenamefont
  {Bellardi}, \citenamefont {Bellazzini}, \citenamefont {Berenji},
  \citenamefont {Bignami}, \citenamefont {Bisello}, \citenamefont {Bissaldi},
  \citenamefont {Blandford}, \citenamefont {Bloom}, \citenamefont {Bogart},
  \citenamefont {Bonamente}, \citenamefont {Bonnell}, \citenamefont {Borgland},
  \citenamefont {Bouvier}, \citenamefont {Bregeon}, \citenamefont {Brez},
  \citenamefont {Brigida}, \citenamefont {Bruel}, \citenamefont {Burnett},
  \citenamefont {Busetto}, \citenamefont {Caliandro}, \citenamefont {Cameron},
  \citenamefont {Caraveo}, \citenamefont {Carius}, \citenamefont {Carlson},
  \citenamefont {Casandjian}, \citenamefont {Cavazzuti}, \citenamefont
  {Ceccanti}, \citenamefont {Cecchi}, \citenamefont {Charles}, \citenamefont
  {Chekhtman}, \citenamefont {Cheung}, \citenamefont {Chiang}, \citenamefont
  {Chipaux}, \citenamefont {Cillis}, \citenamefont {Ciprini}, \citenamefont
  {Claus}, \citenamefont {Cohen-Tanugi}, \citenamefont {Condamoor},
  \citenamefont {Conrad}, \citenamefont {Corbet}, \citenamefont {Corucci},
  \citenamefont {Costamante}, \citenamefont {Cutini}, \citenamefont {Davis},
  \citenamefont {Decotigny}, \citenamefont {DeKlotz}, \citenamefont {Dermer},
  \citenamefont {de~Angelis}, \citenamefont {Digel}, \citenamefont {{do Couto e
  Silva}}, \citenamefont {Drell}, \citenamefont {Dubois}, \citenamefont
  {Dumora}, \citenamefont {Edmonds}, \citenamefont {Fabiani}, \citenamefont
  {Farnier}, \citenamefont {Favuzzi}, \citenamefont {Flath}, \citenamefont
  {Fleury}, \citenamefont {Focke}, \citenamefont {Funk}, \citenamefont {Fusco},
  \citenamefont {Gargano}, \citenamefont {Gasparrini}, \citenamefont {Gehrels},
  \citenamefont {Gentit}, \citenamefont {Germani}, \citenamefont {Giebels},
  \citenamefont {Giglietto}, \citenamefont {Giommi}, \citenamefont {Giordano},
  \citenamefont {Glanzman}, \citenamefont {Godfrey}, \citenamefont {Grenier},
  \citenamefont {Grondin}, \citenamefont {Grove}, \citenamefont {Guillemot},
  \citenamefont {Guiriec}, \citenamefont {Haller}, \citenamefont {Harding},
  \citenamefont {Hart}, \citenamefont {Hays}, \citenamefont {Healey},
  \citenamefont {Hirayama}, \citenamefont {Hjalmarsdotter}, \citenamefont
  {Horn}, \citenamefont {Hughes}, \citenamefont {J\'{o}hannesson},
  \citenamefont {Johansson}, \citenamefont {Johnson}, \citenamefont {Johnson},
  \citenamefont {Johnson}, \citenamefont {Johnson}, \citenamefont {Kamae},
  \citenamefont {Katagiri}, \citenamefont {Kataoka}, \citenamefont {Kavelaars},
  \citenamefont {Kawai}, \citenamefont {Kelly}, \citenamefont {Kerr},
  \citenamefont {Klamra}, \citenamefont {Kn\"{o}dlseder}, \citenamefont
  {Kocian}, \citenamefont {Komin}, \citenamefont {Kuehn}, \citenamefont {Kuss},
  \citenamefont {Landriu}, \citenamefont {Latronico}, \citenamefont {Lee},
  \citenamefont {Lee}, \citenamefont {Lemoine-Goumard}, \citenamefont
  {Lionetto}, \citenamefont {Longo}, \citenamefont {Loparco}, \citenamefont
  {Lott}, \citenamefont {Lovellette}, \citenamefont {Lubrano}, \citenamefont
  {Madejski}, \citenamefont {Makeev}, \citenamefont {Marangelli}, \citenamefont
  {Massai}, \citenamefont {Mazziotta}, \citenamefont {McEnery}, \citenamefont
  {Menon}, \citenamefont {Meurer}, \citenamefont {Michelson}, \citenamefont
  {Minuti}, \citenamefont {Mirizzi}, \citenamefont {Mitthumsiri}, \citenamefont
  {Mizuno}, \citenamefont {Moiseev}, \citenamefont {Monte}, \citenamefont
  {Monzani}, \citenamefont {Moretti}, \citenamefont {Morselli}, \citenamefont
  {Moskalenko}, \citenamefont {Murgia}, \citenamefont {Nakamori}, \citenamefont
  {Nishino}, \citenamefont {Nolan}, \citenamefont {Norris}, \citenamefont
  {Nuss}, \citenamefont {Ohno}, \citenamefont {Ohsugi}, \citenamefont {Omodei},
  \citenamefont {Orlando}, \citenamefont {Ormes}, \citenamefont {Paccagnella},
  \citenamefont {Paneque}, \citenamefont {Panetta}, \citenamefont {Parent},
  \citenamefont {Pearce}, \citenamefont {Pepe}, \citenamefont {Perazzo},
  \citenamefont {Pesce-Rollins}, \citenamefont {Picozza}, \citenamefont
  {Pieri}, \citenamefont {Pinchera}, \citenamefont {Piron}, \citenamefont
  {Porter}, \citenamefont {Poupard}, \citenamefont {Rain\`{o}}, \citenamefont
  {Rando}, \citenamefont {Rapposelli}, \citenamefont {Razzano}, \citenamefont
  {Reimer}, \citenamefont {Reimer}, \citenamefont {Reposeur}, \citenamefont
  {Reyes}, \citenamefont {Ritz}, \citenamefont {Rochester}, \citenamefont
  {Rodriguez}, \citenamefont {Romani}, \citenamefont {Roth}, \citenamefont
  {Russell}, \citenamefont {Ryde}, \citenamefont {Sabatini}, \citenamefont
  {Sadrozinski}, \citenamefont {Sanchez}, \citenamefont {Sander}, \citenamefont
  {Sapozhnikov}, \citenamefont {Parkinson}, \citenamefont {Scargle},
  \citenamefont {Schalk}, \citenamefont {Scolieri}, \citenamefont {Sgr\`{o}},
  \citenamefont {Share}, \citenamefont {Shaw}, \citenamefont {Shimokawabe},
  \citenamefont {Shrader}, \citenamefont {Sierpowska-Bartosik}, \citenamefont
  {Siskind}, \citenamefont {Smith}, \citenamefont {Smith}, \citenamefont
  {Spandre}, \citenamefont {Spinelli}, \citenamefont {Starck}, \citenamefont
  {Stephens}, \citenamefont {Strickman}, \citenamefont {Strong}, \citenamefont
  {Suson}, \citenamefont {Tajima}, \citenamefont {Takahashi}, \citenamefont
  {Takahashi}, \citenamefont {Tanaka}, \citenamefont {Tenze}, \citenamefont
  {Tether}, \citenamefont {Thayer}, \citenamefont {Thayer}, \citenamefont
  {Thompson}, \citenamefont {Tibaldo}, \citenamefont {Tibolla}, \citenamefont
  {Torres}, \citenamefont {Tosti}, \citenamefont {Tramacere}, \citenamefont
  {Turri}, \citenamefont {Usher}, \citenamefont {Vilchez}, \citenamefont
  {Vitale}, \citenamefont {Wang}, \citenamefont {Watters}, \citenamefont
  {Winer}, \citenamefont {Wood}, \citenamefont {Ylinen},\ and\ \citenamefont
  {Ziegler}}]{Atwood2009}%
  \BibitemOpen
  \bibfield  {author} {\bibinfo {author} {\bibfnamefont {W.~B.}\ \bibnamefont
  {Atwood}}, \bibinfo {author} {\bibfnamefont {A.~A.}\ \bibnamefont {Abdo}},
  \bibinfo {author} {\bibfnamefont {M.}~\bibnamefont {Ackermann}}, \bibinfo
  {author} {\bibfnamefont {W.}~\bibnamefont {Althouse}}, \bibinfo {author}
  {\bibfnamefont {B.}~\bibnamefont {Anderson}}, \bibinfo {author}
  {\bibfnamefont {M.}~\bibnamefont {Axelsson}}, \bibinfo {author}
  {\bibfnamefont {L.}~\bibnamefont {Baldini}}, \bibinfo {author} {\bibfnamefont
  {J.}~\bibnamefont {Ballet}}, \bibinfo {author} {\bibfnamefont {D.~L.}\
  \bibnamefont {Band}}, \bibinfo {author} {\bibfnamefont {G.}~\bibnamefont
  {Barbiellini}}, \bibinfo {author} {\bibfnamefont {J.}~\bibnamefont
  {Bartelt}}, \bibinfo {author} {\bibfnamefont {D.}~\bibnamefont {Bastieri}},
  \bibinfo {author} {\bibfnamefont {B.~M.}\ \bibnamefont {Baughman}}, \bibinfo
  {author} {\bibfnamefont {K.}~\bibnamefont {Bechtol}}, \bibinfo {author}
  {\bibfnamefont {D.}~\bibnamefont {B\'{e}d\'{e}r\`{e}de}}, \bibinfo {author}
  {\bibfnamefont {F.}~\bibnamefont {Bellardi}}, \bibinfo {author}
  {\bibfnamefont {R.}~\bibnamefont {Bellazzini}}, \bibinfo {author}
  {\bibfnamefont {B.}~\bibnamefont {Berenji}}, \bibinfo {author} {\bibfnamefont
  {G.~F.}\ \bibnamefont {Bignami}}, \bibinfo {author} {\bibfnamefont
  {D.}~\bibnamefont {Bisello}}, \bibinfo {author} {\bibfnamefont
  {E.}~\bibnamefont {Bissaldi}}, \bibinfo {author} {\bibfnamefont {R.~D.}\
  \bibnamefont {Blandford}}, \bibinfo {author} {\bibfnamefont {E.~D.}\
  \bibnamefont {Bloom}}, \bibinfo {author} {\bibfnamefont {J.~R.}\ \bibnamefont
  {Bogart}}, \bibinfo {author} {\bibfnamefont {E.}~\bibnamefont {Bonamente}},
  \bibinfo {author} {\bibfnamefont {J.}~\bibnamefont {Bonnell}}, \bibinfo
  {author} {\bibfnamefont {A.~W.}\ \bibnamefont {Borgland}}, \bibinfo {author}
  {\bibfnamefont {A.}~\bibnamefont {Bouvier}}, \bibinfo {author} {\bibfnamefont
  {J.}~\bibnamefont {Bregeon}}, \bibinfo {author} {\bibfnamefont
  {A.}~\bibnamefont {Brez}}, \bibinfo {author} {\bibfnamefont {M.}~\bibnamefont
  {Brigida}}, \bibinfo {author} {\bibfnamefont {P.}~\bibnamefont {Bruel}},
  \bibinfo {author} {\bibfnamefont {T.~H.}\ \bibnamefont {Burnett}}, \bibinfo
  {author} {\bibfnamefont {G.}~\bibnamefont {Busetto}}, \bibinfo {author}
  {\bibfnamefont {G.~A.}\ \bibnamefont {Caliandro}}, \bibinfo {author}
  {\bibfnamefont {R.~A.}\ \bibnamefont {Cameron}}, \bibinfo {author}
  {\bibfnamefont {P.~A.}\ \bibnamefont {Caraveo}}, \bibinfo {author}
  {\bibfnamefont {S.}~\bibnamefont {Carius}}, \bibinfo {author} {\bibfnamefont
  {P.}~\bibnamefont {Carlson}}, \bibinfo {author} {\bibfnamefont {J.~M.}\
  \bibnamefont {Casandjian}}, \bibinfo {author} {\bibfnamefont
  {E.}~\bibnamefont {Cavazzuti}}, \bibinfo {author} {\bibfnamefont
  {M.}~\bibnamefont {Ceccanti}}, \bibinfo {author} {\bibfnamefont
  {C.}~\bibnamefont {Cecchi}}, \bibinfo {author} {\bibfnamefont
  {E.}~\bibnamefont {Charles}}, \bibinfo {author} {\bibfnamefont
  {A.}~\bibnamefont {Chekhtman}}, \bibinfo {author} {\bibfnamefont {C.~C.}\
  \bibnamefont {Cheung}}, \bibinfo {author} {\bibfnamefont {J.}~\bibnamefont
  {Chiang}}, \bibinfo {author} {\bibfnamefont {R.}~\bibnamefont {Chipaux}},
  \bibinfo {author} {\bibfnamefont {A.~N.}\ \bibnamefont {Cillis}}, \bibinfo
  {author} {\bibfnamefont {S.}~\bibnamefont {Ciprini}}, \bibinfo {author}
  {\bibfnamefont {R.}~\bibnamefont {Claus}}, \bibinfo {author} {\bibfnamefont
  {J.}~\bibnamefont {Cohen-Tanugi}}, \bibinfo {author} {\bibfnamefont
  {S.}~\bibnamefont {Condamoor}}, \bibinfo {author} {\bibfnamefont
  {J.}~\bibnamefont {Conrad}}, \bibinfo {author} {\bibfnamefont
  {R.}~\bibnamefont {Corbet}}, \bibinfo {author} {\bibfnamefont
  {L.}~\bibnamefont {Corucci}}, \bibinfo {author} {\bibfnamefont
  {L.}~\bibnamefont {Costamante}}, \bibinfo {author} {\bibfnamefont
  {S.}~\bibnamefont {Cutini}}, \bibinfo {author} {\bibfnamefont {D.~S.}\
  \bibnamefont {Davis}}, \bibinfo {author} {\bibfnamefont {D.}~\bibnamefont
  {Decotigny}}, \bibinfo {author} {\bibfnamefont {M.}~\bibnamefont {DeKlotz}},
  \bibinfo {author} {\bibfnamefont {C.~D.}\ \bibnamefont {Dermer}}, \bibinfo
  {author} {\bibfnamefont {A.}~\bibnamefont {de~Angelis}}, \bibinfo {author}
  {\bibfnamefont {S.~W.}\ \bibnamefont {Digel}}, \bibinfo {author}
  {\bibfnamefont {E.}~\bibnamefont {{do Couto e Silva}}}, \bibinfo {author}
  {\bibfnamefont {P.~S.}\ \bibnamefont {Drell}}, \bibinfo {author}
  {\bibfnamefont {R.}~\bibnamefont {Dubois}}, \bibinfo {author} {\bibfnamefont
  {D.}~\bibnamefont {Dumora}}, \bibinfo {author} {\bibfnamefont
  {Y.}~\bibnamefont {Edmonds}}, \bibinfo {author} {\bibfnamefont
  {D.}~\bibnamefont {Fabiani}}, \bibinfo {author} {\bibfnamefont
  {C.}~\bibnamefont {Farnier}}, \bibinfo {author} {\bibfnamefont
  {C.}~\bibnamefont {Favuzzi}}, \bibinfo {author} {\bibfnamefont {D.~L.}\
  \bibnamefont {Flath}}, \bibinfo {author} {\bibfnamefont {P.}~\bibnamefont
  {Fleury}}, \bibinfo {author} {\bibfnamefont {W.~B.}\ \bibnamefont {Focke}},
  \bibinfo {author} {\bibfnamefont {S.}~\bibnamefont {Funk}}, \bibinfo {author}
  {\bibfnamefont {P.}~\bibnamefont {Fusco}}, \bibinfo {author} {\bibfnamefont
  {F.}~\bibnamefont {Gargano}}, \bibinfo {author} {\bibfnamefont
  {D.}~\bibnamefont {Gasparrini}}, \bibinfo {author} {\bibfnamefont
  {N.}~\bibnamefont {Gehrels}}, \bibinfo {author} {\bibfnamefont {F.-X.}\
  \bibnamefont {Gentit}}, \bibinfo {author} {\bibfnamefont {S.}~\bibnamefont
  {Germani}}, \bibinfo {author} {\bibfnamefont {B.}~\bibnamefont {Giebels}},
  \bibinfo {author} {\bibfnamefont {N.}~\bibnamefont {Giglietto}}, \bibinfo
  {author} {\bibfnamefont {P.}~\bibnamefont {Giommi}}, \bibinfo {author}
  {\bibfnamefont {F.}~\bibnamefont {Giordano}}, \bibinfo {author}
  {\bibfnamefont {T.}~\bibnamefont {Glanzman}}, \bibinfo {author}
  {\bibfnamefont {G.}~\bibnamefont {Godfrey}}, \bibinfo {author} {\bibfnamefont
  {I.~A.}\ \bibnamefont {Grenier}}, \bibinfo {author} {\bibfnamefont {M.-H.}\
  \bibnamefont {Grondin}}, \bibinfo {author} {\bibfnamefont {J.~E.}\
  \bibnamefont {Grove}}, \bibinfo {author} {\bibfnamefont {L.}~\bibnamefont
  {Guillemot}}, \bibinfo {author} {\bibfnamefont {S.}~\bibnamefont {Guiriec}},
  \bibinfo {author} {\bibfnamefont {G.}~\bibnamefont {Haller}}, \bibinfo
  {author} {\bibfnamefont {A.~K.}\ \bibnamefont {Harding}}, \bibinfo {author}
  {\bibfnamefont {P.~A.}\ \bibnamefont {Hart}}, \bibinfo {author}
  {\bibfnamefont {E.}~\bibnamefont {Hays}}, \bibinfo {author} {\bibfnamefont
  {S.~E.}\ \bibnamefont {Healey}}, \bibinfo {author} {\bibfnamefont
  {M.}~\bibnamefont {Hirayama}}, \bibinfo {author} {\bibfnamefont
  {L.}~\bibnamefont {Hjalmarsdotter}}, \bibinfo {author} {\bibfnamefont
  {R.}~\bibnamefont {Horn}}, \bibinfo {author} {\bibfnamefont {R.~E.}\
  \bibnamefont {Hughes}}, \bibinfo {author} {\bibfnamefont {G.}~\bibnamefont
  {J\'{o}hannesson}}, \bibinfo {author} {\bibfnamefont {G.}~\bibnamefont
  {Johansson}}, \bibinfo {author} {\bibfnamefont {A.~S.}\ \bibnamefont
  {Johnson}}, \bibinfo {author} {\bibfnamefont {R.~P.}\ \bibnamefont
  {Johnson}}, \bibinfo {author} {\bibfnamefont {T.~J.}\ \bibnamefont
  {Johnson}}, \bibinfo {author} {\bibfnamefont {W.~N.}\ \bibnamefont
  {Johnson}}, \bibinfo {author} {\bibfnamefont {T.}~\bibnamefont {Kamae}},
  \bibinfo {author} {\bibfnamefont {H.}~\bibnamefont {Katagiri}}, \bibinfo
  {author} {\bibfnamefont {J.}~\bibnamefont {Kataoka}}, \bibinfo {author}
  {\bibfnamefont {A.}~\bibnamefont {Kavelaars}}, \bibinfo {author}
  {\bibfnamefont {N.}~\bibnamefont {Kawai}}, \bibinfo {author} {\bibfnamefont
  {H.}~\bibnamefont {Kelly}}, \bibinfo {author} {\bibfnamefont
  {M.}~\bibnamefont {Kerr}}, \bibinfo {author} {\bibfnamefont {W.}~\bibnamefont
  {Klamra}}, \bibinfo {author} {\bibfnamefont {J.}~\bibnamefont
  {Kn\"{o}dlseder}}, \bibinfo {author} {\bibfnamefont {M.~L.}\ \bibnamefont
  {Kocian}}, \bibinfo {author} {\bibfnamefont {N.}~\bibnamefont {Komin}},
  \bibinfo {author} {\bibfnamefont {F.}~\bibnamefont {Kuehn}}, \bibinfo
  {author} {\bibfnamefont {M.}~\bibnamefont {Kuss}}, \bibinfo {author}
  {\bibfnamefont {D.}~\bibnamefont {Landriu}}, \bibinfo {author} {\bibfnamefont
  {L.}~\bibnamefont {Latronico}}, \bibinfo {author} {\bibfnamefont
  {B.}~\bibnamefont {Lee}}, \bibinfo {author} {\bibfnamefont {S.-H.}\
  \bibnamefont {Lee}}, \bibinfo {author} {\bibfnamefont {M.}~\bibnamefont
  {Lemoine-Goumard}}, \bibinfo {author} {\bibfnamefont {A.~M.}\ \bibnamefont
  {Lionetto}}, \bibinfo {author} {\bibfnamefont {F.}~\bibnamefont {Longo}},
  \bibinfo {author} {\bibfnamefont {F.}~\bibnamefont {Loparco}}, \bibinfo
  {author} {\bibfnamefont {B.}~\bibnamefont {Lott}}, \bibinfo {author}
  {\bibfnamefont {M.~N.}\ \bibnamefont {Lovellette}}, \bibinfo {author}
  {\bibfnamefont {P.}~\bibnamefont {Lubrano}}, \bibinfo {author} {\bibfnamefont
  {G.~M.}\ \bibnamefont {Madejski}}, \bibinfo {author} {\bibfnamefont
  {A.}~\bibnamefont {Makeev}}, \bibinfo {author} {\bibfnamefont
  {B.}~\bibnamefont {Marangelli}}, \bibinfo {author} {\bibfnamefont {M.~M.}\
  \bibnamefont {Massai}}, \bibinfo {author} {\bibfnamefont {M.~N.}\
  \bibnamefont {Mazziotta}}, \bibinfo {author} {\bibfnamefont {J.~E.}\
  \bibnamefont {McEnery}}, \bibinfo {author} {\bibfnamefont {N.}~\bibnamefont
  {Menon}}, \bibinfo {author} {\bibfnamefont {C.}~\bibnamefont {Meurer}},
  \bibinfo {author} {\bibfnamefont {P.~F.}\ \bibnamefont {Michelson}}, \bibinfo
  {author} {\bibfnamefont {M.}~\bibnamefont {Minuti}}, \bibinfo {author}
  {\bibfnamefont {N.}~\bibnamefont {Mirizzi}}, \bibinfo {author} {\bibfnamefont
  {W.}~\bibnamefont {Mitthumsiri}}, \bibinfo {author} {\bibfnamefont
  {T.}~\bibnamefont {Mizuno}}, \bibinfo {author} {\bibfnamefont {A.~A.}\
  \bibnamefont {Moiseev}}, \bibinfo {author} {\bibfnamefont {C.}~\bibnamefont
  {Monte}}, \bibinfo {author} {\bibfnamefont {M.~E.}\ \bibnamefont {Monzani}},
  \bibinfo {author} {\bibfnamefont {E.}~\bibnamefont {Moretti}}, \bibinfo
  {author} {\bibfnamefont {A.}~\bibnamefont {Morselli}}, \bibinfo {author}
  {\bibfnamefont {I.~V.}\ \bibnamefont {Moskalenko}}, \bibinfo {author}
  {\bibfnamefont {S.}~\bibnamefont {Murgia}}, \bibinfo {author} {\bibfnamefont
  {T.}~\bibnamefont {Nakamori}}, \bibinfo {author} {\bibfnamefont
  {S.}~\bibnamefont {Nishino}}, \bibinfo {author} {\bibfnamefont {P.~L.}\
  \bibnamefont {Nolan}}, \bibinfo {author} {\bibfnamefont {J.~P.}\ \bibnamefont
  {Norris}}, \bibinfo {author} {\bibfnamefont {E.}~\bibnamefont {Nuss}},
  \bibinfo {author} {\bibfnamefont {M.}~\bibnamefont {Ohno}}, \bibinfo {author}
  {\bibfnamefont {T.}~\bibnamefont {Ohsugi}}, \bibinfo {author} {\bibfnamefont
  {N.}~\bibnamefont {Omodei}}, \bibinfo {author} {\bibfnamefont
  {E.}~\bibnamefont {Orlando}}, \bibinfo {author} {\bibfnamefont {J.~F.}\
  \bibnamefont {Ormes}}, \bibinfo {author} {\bibfnamefont {A.}~\bibnamefont
  {Paccagnella}}, \bibinfo {author} {\bibfnamefont {D.}~\bibnamefont
  {Paneque}}, \bibinfo {author} {\bibfnamefont {J.~H.}\ \bibnamefont
  {Panetta}}, \bibinfo {author} {\bibfnamefont {D.}~\bibnamefont {Parent}},
  \bibinfo {author} {\bibfnamefont {M.}~\bibnamefont {Pearce}}, \bibinfo
  {author} {\bibfnamefont {M.}~\bibnamefont {Pepe}}, \bibinfo {author}
  {\bibfnamefont {A.}~\bibnamefont {Perazzo}}, \bibinfo {author} {\bibfnamefont
  {M.}~\bibnamefont {Pesce-Rollins}}, \bibinfo {author} {\bibfnamefont
  {P.}~\bibnamefont {Picozza}}, \bibinfo {author} {\bibfnamefont
  {L.}~\bibnamefont {Pieri}}, \bibinfo {author} {\bibfnamefont
  {M.}~\bibnamefont {Pinchera}}, \bibinfo {author} {\bibfnamefont
  {F.}~\bibnamefont {Piron}}, \bibinfo {author} {\bibfnamefont {T.~A.}\
  \bibnamefont {Porter}}, \bibinfo {author} {\bibfnamefont {L.}~\bibnamefont
  {Poupard}}, \bibinfo {author} {\bibfnamefont {S.}~\bibnamefont {Rain\`{o}}},
  \bibinfo {author} {\bibfnamefont {R.}~\bibnamefont {Rando}}, \bibinfo
  {author} {\bibfnamefont {E.}~\bibnamefont {Rapposelli}}, \bibinfo {author}
  {\bibfnamefont {M.}~\bibnamefont {Razzano}}, \bibinfo {author} {\bibfnamefont
  {A.}~\bibnamefont {Reimer}}, \bibinfo {author} {\bibfnamefont
  {O.}~\bibnamefont {Reimer}}, \bibinfo {author} {\bibfnamefont
  {T.}~\bibnamefont {Reposeur}}, \bibinfo {author} {\bibfnamefont {L.~C.}\
  \bibnamefont {Reyes}}, \bibinfo {author} {\bibfnamefont {S.}~\bibnamefont
  {Ritz}}, \bibinfo {author} {\bibfnamefont {L.~S.}\ \bibnamefont {Rochester}},
  \bibinfo {author} {\bibfnamefont {A.~Y.}\ \bibnamefont {Rodriguez}}, \bibinfo
  {author} {\bibfnamefont {R.~W.}\ \bibnamefont {Romani}}, \bibinfo {author}
  {\bibfnamefont {M.}~\bibnamefont {Roth}}, \bibinfo {author} {\bibfnamefont
  {J.~J.}\ \bibnamefont {Russell}}, \bibinfo {author} {\bibfnamefont
  {F.}~\bibnamefont {Ryde}}, \bibinfo {author} {\bibfnamefont {S.}~\bibnamefont
  {Sabatini}}, \bibinfo {author} {\bibfnamefont {H.~F.-W.}\ \bibnamefont
  {Sadrozinski}}, \bibinfo {author} {\bibfnamefont {D.}~\bibnamefont
  {Sanchez}}, \bibinfo {author} {\bibfnamefont {A.}~\bibnamefont {Sander}},
  \bibinfo {author} {\bibfnamefont {L.}~\bibnamefont {Sapozhnikov}}, \bibinfo
  {author} {\bibfnamefont {P.~M.~S.}\ \bibnamefont {Parkinson}}, \bibinfo
  {author} {\bibfnamefont {J.~D.}\ \bibnamefont {Scargle}}, \bibinfo {author}
  {\bibfnamefont {T.~L.}\ \bibnamefont {Schalk}}, \bibinfo {author}
  {\bibfnamefont {G.}~\bibnamefont {Scolieri}}, \bibinfo {author}
  {\bibfnamefont {C.}~\bibnamefont {Sgr\`{o}}}, \bibinfo {author}
  {\bibfnamefont {G.~H.}\ \bibnamefont {Share}}, \bibinfo {author}
  {\bibfnamefont {M.}~\bibnamefont {Shaw}}, \bibinfo {author} {\bibfnamefont
  {T.}~\bibnamefont {Shimokawabe}}, \bibinfo {author} {\bibfnamefont
  {C.}~\bibnamefont {Shrader}}, \bibinfo {author} {\bibfnamefont
  {A.}~\bibnamefont {Sierpowska-Bartosik}}, \bibinfo {author} {\bibfnamefont
  {E.~J.}\ \bibnamefont {Siskind}}, \bibinfo {author} {\bibfnamefont {D.~A.}\
  \bibnamefont {Smith}}, \bibinfo {author} {\bibfnamefont {P.~D.}\ \bibnamefont
  {Smith}}, \bibinfo {author} {\bibfnamefont {G.}~\bibnamefont {Spandre}},
  \bibinfo {author} {\bibfnamefont {P.}~\bibnamefont {Spinelli}}, \bibinfo
  {author} {\bibfnamefont {J.-L.}\ \bibnamefont {Starck}}, \bibinfo {author}
  {\bibfnamefont {T.~E.}\ \bibnamefont {Stephens}}, \bibinfo {author}
  {\bibfnamefont {M.~S.}\ \bibnamefont {Strickman}}, \bibinfo {author}
  {\bibfnamefont {A.~W.}\ \bibnamefont {Strong}}, \bibinfo {author}
  {\bibfnamefont {D.~J.}\ \bibnamefont {Suson}}, \bibinfo {author}
  {\bibfnamefont {H.}~\bibnamefont {Tajima}}, \bibinfo {author} {\bibfnamefont
  {H.}~\bibnamefont {Takahashi}}, \bibinfo {author} {\bibfnamefont
  {T.}~\bibnamefont {Takahashi}}, \bibinfo {author} {\bibfnamefont
  {T.}~\bibnamefont {Tanaka}}, \bibinfo {author} {\bibfnamefont
  {A.}~\bibnamefont {Tenze}}, \bibinfo {author} {\bibfnamefont
  {S.}~\bibnamefont {Tether}}, \bibinfo {author} {\bibfnamefont {J.~B.}\
  \bibnamefont {Thayer}}, \bibinfo {author} {\bibfnamefont {J.~G.}\
  \bibnamefont {Thayer}}, \bibinfo {author} {\bibfnamefont {D.~J.}\
  \bibnamefont {Thompson}}, \bibinfo {author} {\bibfnamefont {L.}~\bibnamefont
  {Tibaldo}}, \bibinfo {author} {\bibfnamefont {O.}~\bibnamefont {Tibolla}},
  \bibinfo {author} {\bibfnamefont {D.~F.}\ \bibnamefont {Torres}}, \bibinfo
  {author} {\bibfnamefont {G.}~\bibnamefont {Tosti}}, \bibinfo {author}
  {\bibfnamefont {A.}~\bibnamefont {Tramacere}}, \bibinfo {author}
  {\bibfnamefont {M.}~\bibnamefont {Turri}}, \bibinfo {author} {\bibfnamefont
  {T.~L.}\ \bibnamefont {Usher}}, \bibinfo {author} {\bibfnamefont
  {N.}~\bibnamefont {Vilchez}}, \bibinfo {author} {\bibfnamefont
  {V.}~\bibnamefont {Vitale}}, \bibinfo {author} {\bibfnamefont
  {P.}~\bibnamefont {Wang}}, \bibinfo {author} {\bibfnamefont {K.}~\bibnamefont
  {Watters}}, \bibinfo {author} {\bibfnamefont {B.~L.}\ \bibnamefont {Winer}},
  \bibinfo {author} {\bibfnamefont {K.~S.}\ \bibnamefont {Wood}}, \bibinfo
  {author} {\bibfnamefont {T.}~\bibnamefont {Ylinen}}, \ and\ \bibinfo {author}
  {\bibfnamefont {M.}~\bibnamefont {Ziegler}},\ }\href {\doibase
  10.1088/0004-637X/697/2/1071} {\bibfield  {journal} {\bibinfo  {journal} {The
  Astrophysical Journal}\ }\textbf {\bibinfo {volume} {697}},\ \bibinfo {pages}
  {1071} (\bibinfo {year} {2009})}\BibitemShut {NoStop}%
\bibitem [{\citenamefont {Meegan}\ \emph {et~al.}(2009)\citenamefont {Meegan},
  \citenamefont {Lichti}, \citenamefont {Bhat}, \citenamefont {Bissaldi},
  \citenamefont {Briggs}, \citenamefont {Connaughton}, \citenamefont {Diehl},
  \citenamefont {Fishman}, \citenamefont {Greiner}, \citenamefont {Hoover},
  \citenamefont {van~der Horst}, \citenamefont {von Kienlin}, \citenamefont
  {Kippen}, \citenamefont {Kouveliotou}, \citenamefont {McBreen}, \citenamefont
  {Paciesas}, \citenamefont {Preece}, \citenamefont {Steinle}, \citenamefont
  {Wallace}, \citenamefont {Wilson},\ and\ \citenamefont
  {Wilson-Hodge}}]{Meegan2009}%
  \BibitemOpen
  \bibfield  {author} {\bibinfo {author} {\bibfnamefont {C.}~\bibnamefont
  {Meegan}}, \bibinfo {author} {\bibfnamefont {G.}~\bibnamefont {Lichti}},
  \bibinfo {author} {\bibfnamefont {P.~N.}\ \bibnamefont {Bhat}}, \bibinfo
  {author} {\bibfnamefont {E.}~\bibnamefont {Bissaldi}}, \bibinfo {author}
  {\bibfnamefont {M.~S.}\ \bibnamefont {Briggs}}, \bibinfo {author}
  {\bibfnamefont {V.}~\bibnamefont {Connaughton}}, \bibinfo {author}
  {\bibfnamefont {R.}~\bibnamefont {Diehl}}, \bibinfo {author} {\bibfnamefont
  {G.}~\bibnamefont {Fishman}}, \bibinfo {author} {\bibfnamefont
  {J.}~\bibnamefont {Greiner}}, \bibinfo {author} {\bibfnamefont {A.~S.}\
  \bibnamefont {Hoover}}, \bibinfo {author} {\bibfnamefont {A.~J.}\
  \bibnamefont {van~der Horst}}, \bibinfo {author} {\bibfnamefont
  {A.}~\bibnamefont {von Kienlin}}, \bibinfo {author} {\bibfnamefont {R.~M.}\
  \bibnamefont {Kippen}}, \bibinfo {author} {\bibfnamefont {C.}~\bibnamefont
  {Kouveliotou}}, \bibinfo {author} {\bibfnamefont {S.}~\bibnamefont
  {McBreen}}, \bibinfo {author} {\bibfnamefont {W.~S.}\ \bibnamefont
  {Paciesas}}, \bibinfo {author} {\bibfnamefont {R.}~\bibnamefont {Preece}},
  \bibinfo {author} {\bibfnamefont {H.}~\bibnamefont {Steinle}}, \bibinfo
  {author} {\bibfnamefont {M.~S.}\ \bibnamefont {Wallace}}, \bibinfo {author}
  {\bibfnamefont {R.~B.}\ \bibnamefont {Wilson}}, \ and\ \bibinfo {author}
  {\bibfnamefont {C.}~\bibnamefont {Wilson-Hodge}},\ }\href {\doibase
  10.1088/0004-637X/702/1/791} {\bibfield  {journal} {\bibinfo  {journal} {The
  Astrophysical Journal}\ }\textbf {\bibinfo {volume} {702}},\ \bibinfo {pages}
  {791} (\bibinfo {year} {2009})}\BibitemShut {NoStop}%
\bibitem [{\citenamefont {Tierney}\ \emph {et~al.}(2013)\citenamefont
  {Tierney}, \citenamefont {Briggs}, \citenamefont {Fitzpatrick}, \citenamefont
  {Chaplin}, \citenamefont {Foley}, \citenamefont {McBreen}, \citenamefont
  {Connaughton}, \citenamefont {Xiong}, \citenamefont {Byrne}, \citenamefont
  {Carr}, \citenamefont {Bhat}, \citenamefont {Fishman}, \citenamefont
  {Greiner}, \citenamefont {Kippen}, \citenamefont {Meegan}, \citenamefont
  {Paciesas}, \citenamefont {Preece}, \citenamefont {Kienlin},\ and\
  \citenamefont {Wilson~Hodge}}]{Tierney2013}%
  \BibitemOpen
  \bibfield  {author} {\bibinfo {author} {\bibfnamefont {D.}~\bibnamefont
  {Tierney}}, \bibinfo {author} {\bibfnamefont {M.~S.}\ \bibnamefont {Briggs}},
  \bibinfo {author} {\bibfnamefont {G.}~\bibnamefont {Fitzpatrick}}, \bibinfo
  {author} {\bibfnamefont {V.~L.}\ \bibnamefont {Chaplin}}, \bibinfo {author}
  {\bibfnamefont {S.}~\bibnamefont {Foley}}, \bibinfo {author} {\bibfnamefont
  {S.}~\bibnamefont {McBreen}}, \bibinfo {author} {\bibfnamefont
  {V.}~\bibnamefont {Connaughton}}, \bibinfo {author} {\bibfnamefont
  {S.}~\bibnamefont {Xiong}}, \bibinfo {author} {\bibfnamefont
  {D.}~\bibnamefont {Byrne}}, \bibinfo {author} {\bibfnamefont
  {M.}~\bibnamefont {Carr}}, \bibinfo {author} {\bibfnamefont {P.~N.}\
  \bibnamefont {Bhat}}, \bibinfo {author} {\bibfnamefont {G.~J.}\ \bibnamefont
  {Fishman}}, \bibinfo {author} {\bibfnamefont {J.}~\bibnamefont {Greiner}},
  \bibinfo {author} {\bibfnamefont {R.~M.}\ \bibnamefont {Kippen}}, \bibinfo
  {author} {\bibfnamefont {C.~A.}\ \bibnamefont {Meegan}}, \bibinfo {author}
  {\bibfnamefont {W.~S.}\ \bibnamefont {Paciesas}}, \bibinfo {author}
  {\bibfnamefont {R.~D.}\ \bibnamefont {Preece}}, \bibinfo {author}
  {\bibfnamefont {A.}~\bibnamefont {Kienlin}}, \ and\ \bibinfo {author}
  {\bibfnamefont {C.}~\bibnamefont {Wilson~Hodge}},\ }\href {\doibase
  10.1002/jgra.50580} {\bibfield  {journal} {\bibinfo  {journal} {Journal of
  Geophysical Research: Space Physics}\ } (\bibinfo {year} {2013}),\
  10.1002/jgra.50580}\BibitemShut {NoStop}%
\bibitem [{\citenamefont {Carlson}\ \emph {et~al.}(2007)\citenamefont
  {Carlson}, \citenamefont {Lehtinen},\ and\ \citenamefont
  {Inan}}]{Carlson2007}%
  \BibitemOpen
  \bibfield  {author} {\bibinfo {author} {\bibfnamefont {B.~E.}\ \bibnamefont
  {Carlson}}, \bibinfo {author} {\bibfnamefont {N.~G.}\ \bibnamefont
  {Lehtinen}}, \ and\ \bibinfo {author} {\bibfnamefont {U.~S.}\ \bibnamefont
  {Inan}},\ }\href {\doibase 10.1029/2006GL029229} {\bibfield  {journal}
  {\bibinfo  {journal} {Geophysical Research Letters}\ }\textbf {\bibinfo
  {volume} {34}},\ \bibinfo {pages} {1} (\bibinfo {year} {2007})}\BibitemShut
  {NoStop}%
\bibitem [{\citenamefont {Hazelton}\ \emph {et~al.}(2009)\citenamefont
  {Hazelton}, \citenamefont {Grefenstette}, \citenamefont {Smith},
  \citenamefont {Dwyer}, \citenamefont {Shao}, \citenamefont {Cummer},
  \citenamefont {Chronis}, \citenamefont {Lay},\ and\ \citenamefont
  {Holzworth}}]{Hazelton2009}%
  \BibitemOpen
  \bibfield  {author} {\bibinfo {author} {\bibfnamefont {B.~J.}\ \bibnamefont
  {Hazelton}}, \bibinfo {author} {\bibfnamefont {B.~W.}\ \bibnamefont
  {Grefenstette}}, \bibinfo {author} {\bibfnamefont {D.~M.}\ \bibnamefont
  {Smith}}, \bibinfo {author} {\bibfnamefont {J.~R.}\ \bibnamefont {Dwyer}},
  \bibinfo {author} {\bibfnamefont {X.-M.}\ \bibnamefont {Shao}}, \bibinfo
  {author} {\bibfnamefont {S.~A.}\ \bibnamefont {Cummer}}, \bibinfo {author}
  {\bibfnamefont {T.}~\bibnamefont {Chronis}}, \bibinfo {author} {\bibfnamefont
  {E.~H.}\ \bibnamefont {Lay}}, \ and\ \bibinfo {author} {\bibfnamefont
  {R.~H.}\ \bibnamefont {Holzworth}},\ }\href {\doibase 10.1029/2008GL035906}
  {\bibfield  {journal} {\bibinfo  {journal} {Geophysical Research Letters}\
  }\textbf {\bibinfo {volume} {36}},\ \bibinfo {pages} {L01108} (\bibinfo
  {year} {2009})}\BibitemShut {NoStop}%
\bibitem [{\citenamefont {Gjesteland}\ \emph {et~al.}(2011)\citenamefont
  {Gjesteland}, \citenamefont {\O~stgaard}, \citenamefont {Collier},
  \citenamefont {Carlson}, \citenamefont {Cohen},\ and\ \citenamefont
  {Lehtinen}}]{Gjesteland2011}%
  \BibitemOpen
  \bibfield  {author} {\bibinfo {author} {\bibfnamefont {T.}~\bibnamefont
  {Gjesteland}}, \bibinfo {author} {\bibfnamefont {N.}~\bibnamefont
  {\O~stgaard}}, \bibinfo {author} {\bibfnamefont {A.~B.}\ \bibnamefont
  {Collier}}, \bibinfo {author} {\bibfnamefont {B.~E.}\ \bibnamefont
  {Carlson}}, \bibinfo {author} {\bibfnamefont {M.~B.}\ \bibnamefont {Cohen}},
  \ and\ \bibinfo {author} {\bibfnamefont {N.~G.}\ \bibnamefont {Lehtinen}},\
  }\href {\doibase 10.1029/2011JA016716} {\bibfield  {journal} {\bibinfo
  {journal} {Journal of Geophysical Research}\ }\textbf {\bibinfo {volume}
  {116}},\ \bibinfo {pages} {A11313} (\bibinfo {year} {2011})}\BibitemShut
  {NoStop}%
\bibitem [{\citenamefont {Chaplin}\ \emph {et~al.}(2013)\citenamefont
  {Chaplin}, \citenamefont {Bhat}, \citenamefont {Briggs},\ and\ \citenamefont
  {Connaughton}}]{Chaplin2013}%
  \BibitemOpen
  \bibfield  {author} {\bibinfo {author} {\bibfnamefont {V.}~\bibnamefont
  {Chaplin}}, \bibinfo {author} {\bibfnamefont {N.}~\bibnamefont {Bhat}},
  \bibinfo {author} {\bibfnamefont {M.}~\bibnamefont {Briggs}}, \ and\ \bibinfo
  {author} {\bibfnamefont {V.}~\bibnamefont {Connaughton}},\ }\href
  {http://adsabs.harvard.edu/abs/2013NIMPA.717...21C} {\bibfield  {journal}
  {\bibinfo  {journal} {Nuclear Instruments and Methods in Physics Research A}\
  }\textbf {\bibinfo {volume} {717}},\ \bibinfo {pages} {23} (\bibinfo {year}
  {2013})},\ \Eprint {http://arxiv.org/abs/1211.6592} {arXiv:1211.6592}
  \BibitemShut {NoStop}%
\bibitem [{\citenamefont {Briggs}\ \emph {et~al.}(2010)\citenamefont {Briggs},
  \citenamefont {Fishman}, \citenamefont {Connaughton}, \citenamefont {Bhat},
  \citenamefont {Paciesas}, \citenamefont {Preece}, \citenamefont
  {Wilson-Hodge}, \citenamefont {Chaplin}, \citenamefont {Kippen},
  \citenamefont {von Kienlin}, \citenamefont {Meegan}, \citenamefont
  {Bissaldi}, \citenamefont {Dwyer}, \citenamefont {Smith}, \citenamefont
  {Holzworth}, \citenamefont {Grove},\ and\ \citenamefont
  {Chekhtman}}]{Briggs2010}%
  \BibitemOpen
  \bibfield  {author} {\bibinfo {author} {\bibfnamefont {M.~S.}\ \bibnamefont
  {Briggs}}, \bibinfo {author} {\bibfnamefont {G.~J.}\ \bibnamefont {Fishman}},
  \bibinfo {author} {\bibfnamefont {V.}~\bibnamefont {Connaughton}}, \bibinfo
  {author} {\bibfnamefont {P.~N.}\ \bibnamefont {Bhat}}, \bibinfo {author}
  {\bibfnamefont {W.~S.}\ \bibnamefont {Paciesas}}, \bibinfo {author}
  {\bibfnamefont {R.~D.}\ \bibnamefont {Preece}}, \bibinfo {author}
  {\bibfnamefont {C.}~\bibnamefont {Wilson-Hodge}}, \bibinfo {author}
  {\bibfnamefont {V.~L.}\ \bibnamefont {Chaplin}}, \bibinfo {author}
  {\bibfnamefont {R.~M.}\ \bibnamefont {Kippen}}, \bibinfo {author}
  {\bibfnamefont {A.}~\bibnamefont {von Kienlin}}, \bibinfo {author}
  {\bibfnamefont {C.~A.}\ \bibnamefont {Meegan}}, \bibinfo {author}
  {\bibfnamefont {E.}~\bibnamefont {Bissaldi}}, \bibinfo {author}
  {\bibfnamefont {J.~R.}\ \bibnamefont {Dwyer}}, \bibinfo {author}
  {\bibfnamefont {D.~M.}\ \bibnamefont {Smith}}, \bibinfo {author}
  {\bibfnamefont {R.~H.}\ \bibnamefont {Holzworth}}, \bibinfo {author}
  {\bibfnamefont {J.~E.}\ \bibnamefont {Grove}}, \ and\ \bibinfo {author}
  {\bibfnamefont {A.}~\bibnamefont {Chekhtman}},\ }\href {\doibase
  10.1029/2009JA015242} {\bibfield  {journal} {\bibinfo  {journal} {Journal of
  Geophysical Research}\ }\textbf {\bibinfo {volume} {115}},\ \bibinfo {pages}
  {A07323} (\bibinfo {year} {2010})}\BibitemShut {NoStop}%
\bibitem [{\citenamefont {Dwyer}(2012)}]{Dwyer2012a}%
  \BibitemOpen
  \bibfield  {author} {\bibinfo {author} {\bibfnamefont {J.~R.}\ \bibnamefont
  {Dwyer}},\ }\href {\doibase 10.1029/2011JA017160} {\bibfield  {journal}
  {\bibinfo  {journal} {Journal of Geophysical Research}\ }\textbf {\bibinfo
  {volume} {117}},\ \bibinfo {pages} {A02308} (\bibinfo {year}
  {2012})}\BibitemShut {NoStop}%
\bibitem [{\citenamefont {Celestin}\ \emph {et~al.}(2012)\citenamefont
  {Celestin}, \citenamefont {Xu},\ and\ \citenamefont {Pasko}}]{Celestin2012a}%
  \BibitemOpen
  \bibfield  {author} {\bibinfo {author} {\bibfnamefont {S.}~\bibnamefont
  {Celestin}}, \bibinfo {author} {\bibfnamefont {W.}~\bibnamefont {Xu}}, \ and\
  \bibinfo {author} {\bibfnamefont {V.~P.}\ \bibnamefont {Pasko}},\ }\href
  {\doibase 10.1029/2012JA017535} {\bibfield  {journal} {\bibinfo  {journal}
  {Journal of Geophysical Research}\ }\textbf {\bibinfo {volume} {117}},\
  \bibinfo {pages} {A05315} (\bibinfo {year} {2012})}\BibitemShut {NoStop}%
\bibitem [{\citenamefont {Koshut}\ \emph {et~al.}(1996)\citenamefont {Koshut},
  \citenamefont {Paciesas}, \citenamefont {Kouveliotou}, \citenamefont {van
  Paradijs}, \citenamefont {Pendleton}, \citenamefont {Fishman},\ and\
  \citenamefont {Meegan}}]{Koshut1996}%
  \BibitemOpen
  \bibfield  {author} {\bibinfo {author} {\bibfnamefont {T.~M.}\ \bibnamefont
  {Koshut}}, \bibinfo {author} {\bibfnamefont {W.~S.}\ \bibnamefont
  {Paciesas}}, \bibinfo {author} {\bibfnamefont {C.}~\bibnamefont
  {Kouveliotou}}, \bibinfo {author} {\bibfnamefont {J.}~\bibnamefont {van
  Paradijs}}, \bibinfo {author} {\bibfnamefont {G.~N.}\ \bibnamefont
  {Pendleton}}, \bibinfo {author} {\bibfnamefont {G.~J.}\ \bibnamefont
  {Fishman}}, \ and\ \bibinfo {author} {\bibfnamefont {C.~A.}\ \bibnamefont
  {Meegan}},\ }\href {\doibase 10.1086/177272} {\bibfield  {journal} {\bibinfo
  {journal} {The Astrophysical Journal}\ }\textbf {\bibinfo {volume} {463}},\
  \bibinfo {pages} {570} (\bibinfo {year} {1996})}\BibitemShut {NoStop}%
\bibitem [{\citenamefont {Scargle}\ \emph {et~al.}(2013)\citenamefont
  {Scargle}, \citenamefont {Norris}, \citenamefont {Jackson},\ and\
  \citenamefont {Chiang}}]{Scargle2013}%
  \BibitemOpen
  \bibfield  {author} {\bibinfo {author} {\bibfnamefont {J.~D.}\ \bibnamefont
  {Scargle}}, \bibinfo {author} {\bibfnamefont {J.~P.}\ \bibnamefont {Norris}},
  \bibinfo {author} {\bibfnamefont {B.}~\bibnamefont {Jackson}}, \ and\
  \bibinfo {author} {\bibfnamefont {J.}~\bibnamefont {Chiang}},\ }\href
  {\doibase 10.1088/0004-637X/764/2/167} {\bibfield  {journal} {\bibinfo
  {journal} {The Astrophysical Journal}\ }\textbf {\bibinfo {volume} {764}},\
  \bibinfo {pages} {167} (\bibinfo {year} {2013})}\BibitemShut {NoStop}%
\bibitem [{\citenamefont {Buehler}\ \emph {et~al.}(2012)\citenamefont
  {Buehler}, \citenamefont {Scargle}, \citenamefont {Blandford}, \citenamefont
  {Baldini}, \citenamefont {Baring}, \citenamefont {Belfiore}, \citenamefont
  {Charles}, \citenamefont {Chiang}, \citenamefont {D'Ammando}, \citenamefont
  {Dermer}, \citenamefont {Funk}, \citenamefont {Grove}, \citenamefont
  {Harding}, \citenamefont {Hays}, \citenamefont {Kerr}, \citenamefont
  {Massaro}, \citenamefont {Mazziotta}, \citenamefont {Romani}, \citenamefont
  {{Saz Parkinson}}, \citenamefont {Tennant},\ and\ \citenamefont
  {Weisskopf}}]{Buehler2012}%
  \BibitemOpen
  \bibfield  {author} {\bibinfo {author} {\bibfnamefont {R.}~\bibnamefont
  {Buehler}}, \bibinfo {author} {\bibfnamefont {J.~D.}\ \bibnamefont
  {Scargle}}, \bibinfo {author} {\bibfnamefont {R.~D.}\ \bibnamefont
  {Blandford}}, \bibinfo {author} {\bibfnamefont {L.}~\bibnamefont {Baldini}},
  \bibinfo {author} {\bibfnamefont {M.~G.}\ \bibnamefont {Baring}}, \bibinfo
  {author} {\bibfnamefont {A.}~\bibnamefont {Belfiore}}, \bibinfo {author}
  {\bibfnamefont {E.}~\bibnamefont {Charles}}, \bibinfo {author} {\bibfnamefont
  {J.}~\bibnamefont {Chiang}}, \bibinfo {author} {\bibfnamefont
  {F.}~\bibnamefont {D'Ammando}}, \bibinfo {author} {\bibfnamefont {C.~D.}\
  \bibnamefont {Dermer}}, \bibinfo {author} {\bibfnamefont {S.}~\bibnamefont
  {Funk}}, \bibinfo {author} {\bibfnamefont {J.~E.}\ \bibnamefont {Grove}},
  \bibinfo {author} {\bibfnamefont {A.~K.}\ \bibnamefont {Harding}}, \bibinfo
  {author} {\bibfnamefont {E.}~\bibnamefont {Hays}}, \bibinfo {author}
  {\bibfnamefont {M.}~\bibnamefont {Kerr}}, \bibinfo {author} {\bibfnamefont
  {F.}~\bibnamefont {Massaro}}, \bibinfo {author} {\bibfnamefont {M.~N.}\
  \bibnamefont {Mazziotta}}, \bibinfo {author} {\bibfnamefont {R.~W.}\
  \bibnamefont {Romani}}, \bibinfo {author} {\bibfnamefont {P.~M.}\
  \bibnamefont {{Saz Parkinson}}}, \bibinfo {author} {\bibfnamefont {A.~F.}\
  \bibnamefont {Tennant}}, \ and\ \bibinfo {author} {\bibfnamefont {M.~C.}\
  \bibnamefont {Weisskopf}},\ }\href {\doibase 10.1088/0004-637X/749/1/26}
  {\bibfield  {journal} {\bibinfo  {journal} {The Astrophysical Journal}\
  }\textbf {\bibinfo {volume} {749}},\ \bibinfo {pages} {26} (\bibinfo {year}
  {2012})}\BibitemShut {NoStop}%
\end{thebibliography}%

\end{document}